\documentclass{amsart}

\usepackage{amsthm,amsmath,amssymb,url,hyperref,graphicx,color,fullpage}
\usepackage{stmaryrd} 

\newtheorem{theorem}{Theorem}[section]

\newtheorem{proposition}[theorem]{Proposition}
\newtheorem{lemma}[theorem]{Lemma}

\newtheorem{corollary}[theorem]{Corollary}

\theoremstyle{remark}

\newtheorem{remark}{Remark}

\newcommand{\BR}{{\mathbb R}}

\newcommand{\BE}{{\mathbb E}}
\newcommand{\BP}{{\mathbb P}}

\newcommand{\CL}{{\mathcal L}}

\newcommand{\CE}{{\mathcal E}}
\newcommand{\CP}{{\mathcal P}}

\newcommand{\one}{{\mathbf 1}}

\newcommand{\ov}{\overline}

\newcommand{\del}{\partial}

\providecommand{\norm}[1]{\left\lVert#1\right\rVert}

\newcommand{\bin}[1]{\llbracket#1\rrbracket}

\newcommand{\bid}{\text{bid}}
\newcommand{\ask}{\text{ask}}

\title[Markovian Limit Order Book]{A Markov model of a limit order book:  thresholds, recurrence, 
and trading strategies}

\author[Kelly and Yudovina]{Frank Kelly \and Elena Yudovina}

\thanks{The second author's research was partially supported by NSF Graduate Research Fellowship and NSF grant DMS-1204311.}

\address{
Frank Kelly\\
Statistics Laboratory, Centre for Mathematical Sciences\\
University of Cambridge\\
Wilberforce Rd\\
Cambridge CB3 0WA\\
E-mail: {\tt fpk1@cam.ac.uk}\\
Elena Yudovina\\
Department of Mathematics, University of Minnesota\\
127 Vincent Hall\\
206 Church St. S.E.\\
Minneapolis, MN 55455\\
E-mail: {\tt eyudovin@umn.edu}}

\begin{document}

\begin{abstract}%
We analyze a tractable model of a limit order book on short
time scales, where the dynamics are driven by stochastic fluctuations
between supply and demand.
We establish the existence of a limiting distribution for the
highest bid, and for the lowest ask, where the limiting distributions are
confined between two thresholds. We make extensive use of fluid limits in
order to establish recurrence properties of the model. We use the model to
analyze various high-frequency trading strategies, and comment on 
the Nash equilibria that emerge between high-frequency traders when a market in continuous time is replaced by frequent batch auctions.
\end{abstract}%

\maketitle

\section{Introduction.} \label{sec:intro} 

A limit order book (LOB) is a trading mechanism for a single-commodity market. The mechanism is of significant interest to economists as a model of price formation. It is also used in many financial markets, and has generated extensive research, both empirical and theoretical: for a recent survey, see \cite{GPSMFD}.

The detailed historic data from LOBs in financial markets has encouraged models able to replicate the observed statistical properties of these markets. Unfortunately, the added complexity usually makes the models less analytically 
tractable and, with relatively few exceptions, such models  
are explored by simulation or numerical methods. Our aim in this paper is
to analyze a simple and tractable model of a LOB, first introduced by
\cite{Stigler} and independently  by \cite{Luckock}
and by \cite{Plackova}.  The basic form of the model 
explicitly excludes 
a number of significant features of
real-world markets. Nevertheless we shall see that, from the model, 
several non-trivial and insightful results can be obtained on 
the structure of high-frequency
trading strategies.
Further, the model has a natural interpretation for a competitive and highly
traded market on short time-scales, where the excluded features may be less
significant. 
We believe the model may be helpful in discussions of market design, 
and as an
illustration we use the model to comment on the Nash equilibria 
that emerge between high-frequency traders when a market in 
continuous time is replaced by frequent
batch auctions. 





To motivate the model consider
a market with only two classes of participant. Firstly, 
long-term investors who place orders for reasons exogenous
to the model,\footnote{For example, to manage their portfolios. Investors
may differ in their preferences and in their valuations, even given the
same information, which creates potential gains from trade.} who view
the market as effectively efficient for their purposes, and who
do not shade their orders strategically. Temporary 
imbalances between supply and demand from such long-term investors 
will cause prices to fluctuate even
in the absence of any new information becoming available concerning the
fundamentals of the underlying asset. Our second 
class of participant, high-frequency
traders, attempt to benefit from these price fluctuations
by providing liquidity between the long-term investors. In practice
we should expect a spectrum of behavior between these
two extremes. 
The extreme case, with 
just long-term investors and high-frequency traders, 
is clearly a caricature,
but we shall see that it does allow us to analyze 
various high-frequency
trading strategies (for example market-making, sniping and mixtures of
these) and the Nash equilibria between them.

We next describe the model of a LOB for an example involving
long-term investors only, and outline 
our results for this
example. A \emph{bid} is an order to buy one unit, and an \emph{ask} is an
order to sell one unit. Each order has associated with it a \emph{price}, a
real number. Suppose that bids and asks arrive as independent Poisson
processes of unit rate
and that the prices associated with bids, respectively asks, are independent identically distributed random variables with density $f_b(x)$, respectively $f_a(x)$. An arriving bid is either added to the LOB, if it is lower than all asks present in the LOB, or it is matched to the lowest ask and both depart. Similarly an arriving ask is either added to the LOB, if it is higher than all bids present in the LOB, or it is matched to the highest bid and both depart. The LOB at time $t$ is thus the set of bids and asks (with their prices), and our assumptions imply the LOB is a Markov process.

For this model we show that there exists a threshold $\kappa_b$ with the
following properties: for any $x<\kappa_b$ there is a finite time after
which no arriving bids less than $x$ are ever matched; and for any
$x>\kappa_b$ the event that there are no bids greater than $x$ in the LOB
is recurrent. Similarly, with directions of inequality reversed, there
exists a corresponding threshold $\kappa_a$ for asks. Further there is a
density $\pi_a(x)$, respectively $\pi_b(x)$, supported on $(\kappa_b, \kappa_a)$ giving the 
limiting distribution of the lowest ask, respectively highest bid, in the LOB. The densities $\pi_a, \pi_b$ solve the equations
\begin{subequations}\label{eq: integral_intro}
\begin{equation} \label{eq: integral_intro_1}
f_b(x) \int_x^{\kappa_a} \pi_a(y) dy = \pi_b(x)\int_{-\infty}^x f_a(y) dy 
\end{equation}
\begin{equation} \label{eq: integral_intro_2}
f_a(x) \int_{\kappa_b}^x \pi_b(y) dy = \pi_a(x)\int_x^{\infty}  f_b(y) dy. 
\end{equation}
\end{subequations}
As a specific example, 
if $f_a(x)=f_b(x)=1, x \in (0,1)$, then $\kappa_a = \kappa, \kappa_b = 1- \kappa$, $\pi_a(x) = \pi_b(1-x)$, and 
\begin{equation} \label{eq: soln}
\pi_b(x) = (1-\kappa) \left(\frac1x +\log\left(\frac{1-x}{x}\right) \right), \quad x \in (\kappa, 1- \kappa) 
\end{equation}
where the value of $\kappa$ is given as follows. Let $w$ be the unique solution of $ w e^w = e^{-1}$: then $ w \approx 0.278$ and $\kappa = w/(w+1) \approx 0.218$. Observe that any example with $f_a=f_b$ can be reduced to this example by a monotone transformation of the price axis.



The existence of thresholds with the claimed properties is a relatively straightforward result, using Kolmogorov's 0--1 law. In order to make the claimed distributional result precise the major challenge is to establish positive recurrence of certain \emph{binned} models: such models arise naturally where, for example, prices are recorded to only a finite number of decimal places. Given a sufficiently strong notion of recurrence the intuition behind equations~\eqref{eq: integral_intro} is straightforward: in equilibrium the right-hand side of equation~\eqref{eq: integral_intro_1} is the probability flux that the highest bid in the LOB is at $x$ and that it is matched by an arriving ask with a price less than $x$, and the left-hand side is the probability flux that the lowest ask in the LOB is more than $x$ and that an arriving bid enters the LOB at price $x$; these must balance, and a similar argument for the lowest ask leads to equation~\eqref{eq: integral_intro_2}. To establish positive recurrence of the binned models we make extensive use of fluid limits (see \cite{Bramson}), an important technique in the study of queueing networks.

The orders we have described so far are called \emph{limit orders} to
distinguish them from \emph{market orders} which request to be fulfilled
immediately at the best available price. Market orders are straightforward
to include in the model: in the specific 
example just described we simply associate
a price $1$ or $0$ with a market bid or market ask respectively. As the
proportion of market bids increases towards a critical threshold, $ w
\approx 0.278$ in the above example, the support of the 
limiting distributions $\pi_a, \pi_b$ increases to approach the entire
interval $(0,1)$: above the threshold the model predicts recurring periods
of time when there will exist either no highest bid or no lowest ask in the
LOB. This conclusion necessarily 
holds, with the same critical threshold $w$, for any example with $f_a=f_b$.

A LOB is a form of two-sided queue, the study of which dates
at least to the early paper of \cite{KEN},
 who modeled a taxi-stand with arrivals of both taxis and travellers as a symmetric random walk. Recent theoretical advances involve servers and customers with varying types and constraints on feasible matchings between servers and customers, with applications ranging from large-scale call centres to national waiting lists for organ transplants (cf. \cite{AW, SY, ZCW}). Our interest in models of LOBs is in part due to the simplicity of the matchings in this particular application: types, as real variables, are totally ordered and so when an arriving order can be matched the match is uniquely defined.

Next we comment on several 
important features of real-world markets that are missing from the above basic
model of a LOB. We assume that orders (from investors) 
are never cancelled and that the arrival streams of orders, with their prices, are 
not dependent on the state of the LOB. These assumptions might be natural
for orders from our long-term investors who view the market as
effectively efficient for their purposes. These assumptions,
and the related assumption of stationarity of the arrival streams, may 
also be natural for a high-volume market 
where there may be a substantial amount of trading 
activity even over time periods where 
no new information becomes available 
concerning the fundamentals of the underlying asset. 
Mathematically the model may then be viewed as assuming a 
separation between the time-scale of trading, represented
in the model, and a longer time-scale on which fundamentals
change. 

The assumption that all orders are for a single unit is
important mathematically for the derivation of
equations~\eqref{eq: integral_intro}; economically, it corresponds
to an assumption of a competitive market
where an investor does not need to think about the impact
of her order size on the market. 
We note that a
long-term investor placing a large order may attempt to be passive in her
execution, so as not to move the price against her, by spreading the order
in line with volume in the market; see \cite{ELO}. The natural
question then becomes over how long  the order is spread,
and the model can give insight here. 
We note, however, that 
our assumption of a separation between the time-scale of
trading and the timescale
on which fundamentals change, modeled by our assumption of stationarity of
arrival streams, may no longer be tenable when the time taken 
by a large investor to complete an order increases. 
In markets with
 a relatively small set of participants with large
orders other approaches may be necessary; see  \cite{DZ} for 
a discussion of trading protocols that complement limit order books
for large strategic investors.

Markets may contain traders other than long-term investors, and there is
currently considerable interest in the effect of high-frequency trading on
LOBs. Importantly, many high-frequency trading strategies are
straightforward to represent within the 
model, since traders who can react
immediately to an order entering the LOB may leave the Markov structure
intact. Consider first the following \emph{sniping} strategy for a single
high-frequency trader: she immediately buys every bid that joins the LOB at
price above $q$ and every ask that joins the LOB at price below $p$, where
$p$ and $q$ are chosen to balance the rates of these purchases. This model
fits straightforwardly within our framework, and we show how to calculate
the optimal values, for the high-frequency trader, of the constants $p$ and
$q$. A single trader might instead behave as a~\emph{market maker} and
place an infinite number of bid, respectively ask, orders at $p$,
respectively $q$, where $\kappa_b < p<q < \kappa_a $. We are again able to
analyze this case. The optimal profit rate under the sniping strategy
may beat that under the market making strategy: it does so for the specific
example above where  $f_a(x)=f_b(x)=1, x \in (0,1)$, describes the order flow from long-term investors. But a third strategy, which combines market making and sniping, will generally beat both the individual strategies.

The model also allows us to readily explore the
equilibria that emerge when there are
multiple high-frequency traders competing using 
market making or sniping strategies. 
There has been considerable discussion recently of the effects
of competition between multiple high-frequency traders, and of proposals
aimed to slow down markets. A key issue is that high-frequency traders may
wastefully compete on the speed with which they can snipe an order, 
and as a regulatory response 
Budish et al.~\cite{BCS, BCSAER} propose replacing a
market continuous in time with frequent batch auctions, held perhaps
several times a second. We consider
Nash equilibria in continuous and batch markets when
there are multiple high-frequency traders competing using mixtures of
market making and sniping. Competition between market making traders
reduces the bid-ask spread 
and the traders' profit rate, 
and does so whether the market is continuous or batch. 
Competition between sniping traders in a batch market
results in a Nash equilibrium with 
traders sniping bids above, respectively asks below, a central price;
the traders' profit rate is slightly less in a batch market than
a continuous market.

Competition between sniping strategies 
produces a large number of cancelled orders 
since if a strategy's attempt to snipe an arriving order is not 
successful then the strategy immediately 
cancels its own order. 
A notable feature of data from real LOBs, that a substantial
proportion of orders are immediately cancelled~\cite{GPSMFD}, 
thus emerges as a deduction
from, rather than an assumption of, the model. 


A discrete version of the model was first proposed by \cite{Stigler} 
in his pioneering work on regulation of securities markets, 
and the model was independently introduced 
by \cite{Luckock}
and by \cite{Plackova}.  Taking stationarity as an 
assumption, 
\cite{Luckock} provided an extensive analysis of the model; our 
equations~\eqref{eq: integral_intro} can be deduced
from \cite[Proposition 1]{Luckock}, assuming 
steady-state behavior, 
by setting time derivatives to zero. Our contribution is to
establish the existence of the thresholds $\kappa_a, \kappa_b$
and to prove a sufficiently strong notion of recurrence to justify
the intuition behind equations~\eqref{eq: integral_intro}.

Previous research similar in mathematical framework to that
reported here is by Cont and coauthors~\cite{CST, Cont_deLarrard},
by Simatos and coauthors~\cite{SIM, LRS} building on 
Lakner et al.~\cite{LRSt}, and by Toke~\cite{TOK}: as we do, these 
authors 
describe LOBs as Markovian systems of interacting queues and are able to obtain analytical expressions for various quantities of interest. In the
models of~\cite{CST, Cont_deLarrard, LRSt, SIM, LRS}
the arrival rates of orders at any given price depends on how far
the price is above or below the current best ask or bid price;
the models of~\cite{LRSt, SIM, LRS, TOK} are one-sided in that all bids
are limit orders and all asks are market orders.  Gao et
al.~\cite{GDDD} study the temporal evolution of the the shape of a LOB in
the model of~\cite{CST}, under a scaling limit. Maglaras et al.~\cite{MMZ}
study a fragmented one-sided market in which traders may route their orders
to one of several exchanges.
The work of Lachappelle et al.~\cite{LLLL}, building on Ro\c{s}u~\cite{Rosu}, uses a different mathematical framework, that of a mean field game, but shares with our approach some important features. In particular, these authors distinguish between institutional investors whose decisions are independent of the immediate state of the LOB
and high-frequency traders who trade as a consequence of the immediate
state of the LOB. The models of both~\cite{Cont_deLarrard} and~\cite{LLLL}
keep detailed information on queue sizes only at the best bid and best ask
prices; \cite{CST} shares with our approach a Markov description of the entire LOB.

In much of the market microstructure literature features of LOBs, such as
large bid-ask spreads, are explained as a consequence of participants
protecting themselves from others with superior information. While this is 
clearly an important aspect of real-world markets we note that
such features may also arise from simpler models. 
The driving force for the dynamics of the LOB in our approach, as in~\cite{LLLL, Rosu}, is not asymmetric information but 
stochastic fluctuations between supply and demand.

The organisation of the paper is as follows. In Section~\ref{sec:model} 
we describe precisely the model and our main results.
Section~\ref{sec:scaffold} develops the scaffolding necessary for the
proofs, which are given in Section~\ref{sec:proofs}. In
Section~\ref{sec:applications} we describe some applications of our
results: this section contains our discussion of market orders,
 and of 
high-frequency trading strategies and Nash equilibria.

\section{Model and results.}\label{sec:model}\label{sec:results}


The state of the LOB at time $t$ is a pair $(B_t, A_t)$ of (possibly
infinite) counting measures on $\BR$; $B_t$ represents the prices of queued (not yet executed) bid orders, and $A_t$ represents the
prices of queued asks. New orders arrive as a labeled point process;
the label records the type of order (bid or ask) and the price. Without loss of generality, we assume that the price axis has been continuously reparametrized so that all prices fall in the interval $(0,1)$ (or, occasionally, $[0,1]$).

Orders depart from the queue when an arriving order ``matches'' one of the
orders already in the book. We shall need several notions of what it means
for two prices to match, and to capture this we introduce
a \emph{price equivalence function}, that is a
nondecreasing, not necessarily continuous, function $\CP: [0,1] \to [0,1]$.
A bid-ask
pair is \emph{compatible} if $\CP(\bid) \geq \CP(\ask)$.%
We shall primarily
consider two types of price equivalence function: $\CP(x) = x$, and the function
that partitions all prices into $n$ pricing bins. We will refer to the 
latter case, where the image of $\CP$ is a finite set,  
as the \emph{binned model}.  Note that the same price equivalence function is applied to the prices of all the orders, and compatibility of bid-ask pairs is unchanged under any strictly increasing transformation of the equivalence function.

We are now ready to formally define the evolution limit order book $\CL_t$.

\noindent{\bf Initial state:} Initially, there should be no compatible bid--ask pairs in the book. Equivalently, the initial state $(B_0,A_0)$ satisfies
\[
B_0[x,1) \cdot A_0(0,y] = 0
\quad \text{ if } \quad
\CP(y) \leq \CP(x).
\]
Most of the time we assume that the total number of orders in the book is finite; we relax this assumption in Section~\ref{sec:applications}, where we allow an infinite number of orders to be placed at a single price, and otherwise the book is finite.

\noindent{\bf Order arrival process:} New orders arrive as a Poisson process with iid labels designating the type and price of the order. Unless specified otherwise, we assume that $\BP(\text{bid}) = \BP(\text{ask}) = 1/2$.%
\footnote{The Poisson structure is not important to the book, because all that matters is the sequence of order arrivals. Unequal rates of arrival for bids and asks are considered in Section~\ref{sec:differing}.}
We assume the labels of orders are independent and identically distributed, and in particular independent of the state of the book, but the distributions of prices may depend on type. We let $F_a$ be the CDF of prices of arriving asks, and $F_b$ be the CDF of prices of arriving bids. We will often assume that the distributions of the prices of arriving orders have densities $f_a$ and $f_b$ respectively; this entails no loss of generality, because the LOB evolution is defined by the combination of the arriving price distributions and a price equivalence function, and thus we can always assume that the arriving orders have densities and only become discontinuous after being put through the price equivalence function.

\noindent{\bf Change at order arrival:} We do not allow cancellations in
the model (until Section~\ref{sec:applications}), so all changes to the state occur at the time of an order
arrival. Suppose at time $t$ a bid at price $p$ arrives. If there is a
matching ask in the book, i.e. 
if $A_{t-}(0,y] > 0$ for some $y$ such that $\CP(y) \leq \CP(x)$, then nothing happens to the bids in the book ($B_t = B_{t-}$), and the lowest ask departs: $A_t = A_{t-} - \delta_q$, where $q = \min\{x: A_{t-}\{x\} > 0\}$%
\footnote{The minimum exists when the initial state of the book is finite, since only finitely many orders are present in the book.}
. If there are no matching asks in the book, the bid joins the book: $B_t = B_{t-} + \delta_p$ and $A_t = A_{t-}$. The situation is symmetric if the arriving order is an ask at price $q$: if there is a matching bid, the two orders depart (so $A_t = A_{t-}$ and $B_t = B_{t-} - \delta_p$ where $p = \max\{x: B_{t-}\{x\} > 0\}$), and if there are no matching bids, then the ask joins the book ($B_t = B_{t-}$ and $A_t = A_{t-} + \delta_q$).


\vspace{\baselineskip}

We will be keeping track of the highest (price of a) bid $\beta_t$ and lowest (price of an) ask $\alpha_t$ in the book at time $t$. If an order departs the book at time $t$, it must be at price $\beta_{t-}$ (if a bid) or $\alpha_{t-}$ (if an ask). We allow $B_0\{x\} = \infty$ or $A_0\{y\} = \infty$; if this is the case, then no bids left of $x$, and no asks right of $y$, will ever depart the limit order book, since they will never be the highest bid (respectively lowest ask).


Below, we will refer to continuous and discretized models of LOBs. A continuous LOB is one where the order price densities $f_a$ and $f_b$ (exist and) are bounded above and below, and the price equivalence function is $\CP(x) = x$. Discretized models will use some binned price equivalence function, and will sometimes (but not always) assume that all bins receive a positive proportion of the orders of each type.

For a discretized, binned LOB, we will use notation $\bin{x}$ to denote the index of the bin containing $x$; $\bin{x}$ is a positive integer ranging from 1 to $N$ for some $N > 0$.

We now present the main results concerning the model. The first
result, Theorem~\ref{thm: threshold}, 
establishes a transition at threshold values $\kappa_b$ and
$\kappa_a$. Eventually bids arriving below $\kappa_b$, and asks arriving
above $\kappa_a$, will never be executed; whereas all bids arriving
above $\kappa_b$, and all asks arriving below $\kappa_a$, will be executed. 
The second result, Theorem~\ref{thm: distro}, presents the distribution of the rightmost bid and leftmost ask.

\begin{theorem}[Thresholds]\label{thm: threshold}
There exist prices $\kappa_b$ and $\kappa_a$ with the following properties:
\begin{enumerate}
\item For any $\epsilon > 0$ there exists, almost surely, a (random) time $T_0 < \infty$ 
such that $\beta_t > \kappa_b - \epsilon$ and $\alpha_t < \kappa_a + \epsilon$ for all $t \geq T_0$.
\item For any $\epsilon > 0$, infinitely often there will be no orders with prices in $(\kappa_b+\epsilon,\kappa_a-\epsilon)$.
\item Let $x > \kappa_b + \epsilon$ and $y < \kappa_a-\epsilon$ for some $\epsilon > 0$. Consider the LOB started with infinitely many bids at $x$, infinitely many asks at $y$, and finitely many orders in between. 
The evolution of the orders at prices in the interval $(x,y)$ 
is a positive (Harris) recurrent Markov process, with finite expected time until there are 
no orders in the interval.
\end{enumerate}
\end{theorem}

The fact that there will infinitely often be no bids above $x$, and no asks
below $y$, is a consequence of Kolmogorov's 0--1 law; the challenge is to
show that there will \emph{simultaneously} be neither bids nor asks 
in the interval $(x,y)$. 
In fact, we shall need to prove this 
part of Theorem~\ref{thm: threshold} and Theorem~\ref{thm: distro} below in tandem.

\begin{theorem}[Distribution of the highest bid]\label{thm: distro}
Consider a continuous LOB; that is, $\CP(x) = x$, and the densities $f_b$ and $f_a$ are bounded above and below. Then
\begin{enumerate}
\item The limiting distributions of the highest bid and lowest ask have densities, denoted $\pi_b$ and $\pi_a$; let $\varpi_b = \pi_b/f_b$ and $\varpi_a = \pi_a / f_a$.
\item The thresholds satisfy $0 < \kappa_b < \kappa_a < 1$, and also $F_b(\kappa_b) = 1-F_a(\kappa_a)$.
\item The distribution of the highest bid is such that $\varpi_b$ 
is the unique solution to the 
ordinary differential equation
\[
\left(-\frac{f_a(x)}{1-F_b(x)}(F_a(x)\varpi_b(x))'\right)' = \varpi_b(x)f_b(x)
\]
with initial conditions
\[
(F_a(x)\varpi_b(x)) \vert_{x = \kappa_b} = 1, \qquad (F_a(x)\varpi_b(x))'\vert_{x=\kappa_b} = 0
\]
and the additional constraint $\varpi_b(x) \to 0$ as $x \uparrow \kappa_a$.
The distribution of the lowest ask is determined by a similar ODE.
\end{enumerate}
\end{theorem}
\begin{corollary}[Uniform arrivals]\label{coroll: uniform}
Suppose that $\CP(x) = x$ and the arrival price distribution is uniform on
$(0,1)$ for both bids and asks. Then $\kappa_b = \kappa \approx 0.218$ 
is given by
$\kappa = w/(w+1)$ where $we^w=e^{-1}$. The limiting 
density of the highest bid is supported on $(\kappa, 1-\kappa)$, and is given by
\[
\varpi_b(x) = \one_{(\kappa,1-\kappa)}(1-\kappa) \left(\frac1x +\log
\left(\frac{1-x}{x}\right) \right)
\]
and the limiting density of the lowest ask is $\varpi_a(x) =
\varpi_b(1-x)$.
\end{corollary}

\begin{remark}[Absolute continuity]
We can replace conditions on the densities $f_a$ and $f_b$ by the requirement that $dF_a/dF_b$ be bounded above and below; however, it is more natural to state the result of Theorem~\ref{thm: distro} in terms of densities. The boundedness requirement avoids the trivial counterexamples $f_b = 2\one_{[0,1/2)}$, $f_a = 2\one_{(1/2,1]}$ (nonoverlapping supports, no orders leave) or $f_a = 2\one_{[0,1/2)}$, $f_b = 2\one_{(1/2,1]}$ (nonoverlapping supports, no threshold). 

Through a reparametrization of the price axis, Corollary~\ref{coroll: uniform} 
covers all cases where
arriving bid and ask prices have identical densities.  We
describe some other analytically tractable applications of Theorem~\ref{thm:
distro} in Section~\ref{sec:applications}. We shall also, in
Section~\ref{sec:applications}, extend
the analysis to deal with some examples where the
supports of the bid and ask price distributions do not coincide.

\end{remark}



\begin{remark}

The form of the limiting density appearing in 
Corollary~\ref{coroll: uniform} 
can be deduced from equations (63)--(64) of
\cite[Section 3]{Luckock} after applying a coordinate transformation to
convert between $[0,\infty)$ and $[0,1)$.

\end{remark}

\begin{figure}
  \begin{center}
    {\includegraphics[width=0.5\textwidth]{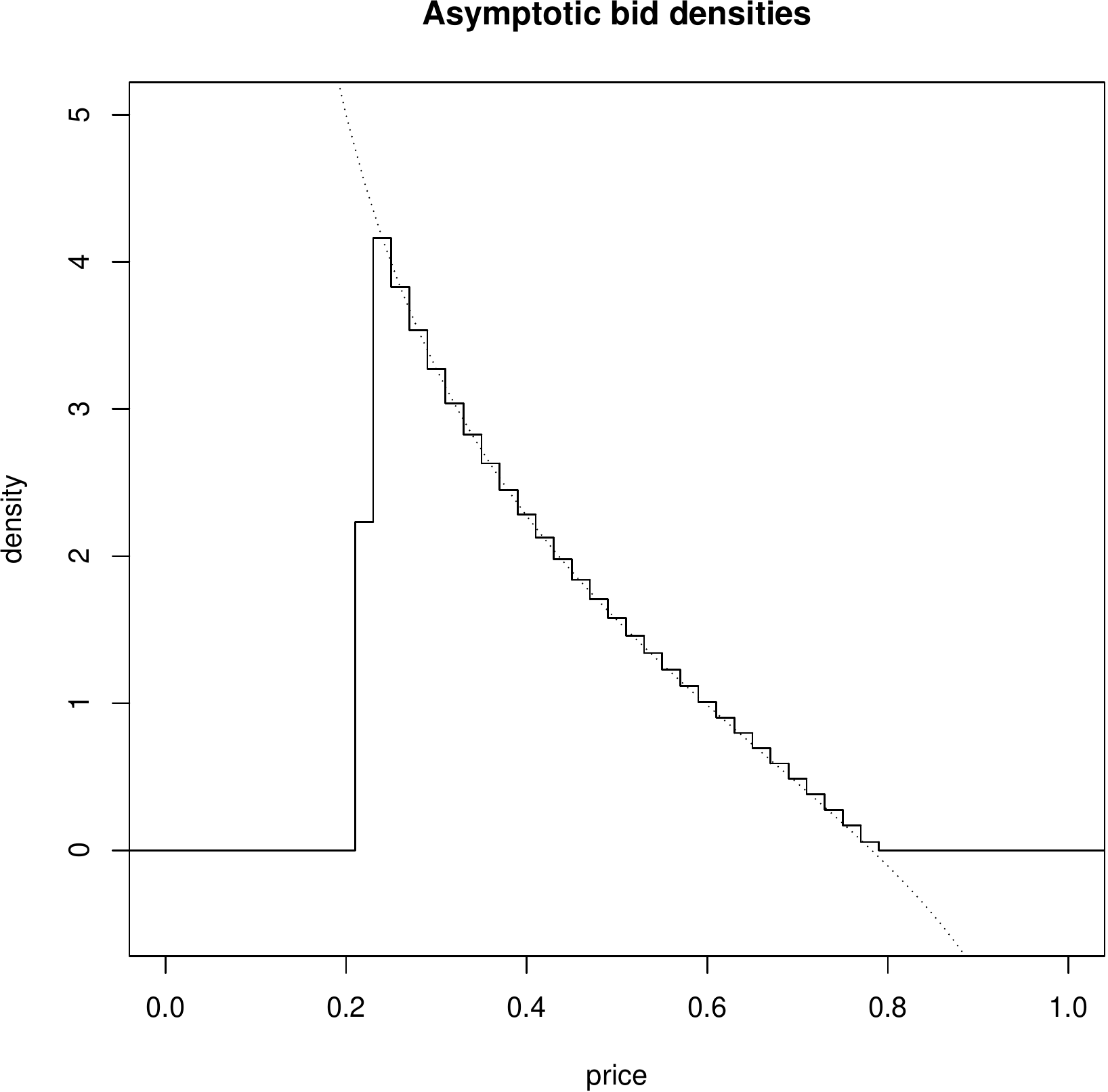}}
  \end{center}
  \caption{Limiting density of the highest bid for the binned LOB with 50 bins, and limiting density for a continuous LOB (dotted line). Note the ``shoulder'' bin in the binned model: the threshold in the continuous LOB lies in the interior of this bin.}
  \label{fig:density}
\end{figure}

In Figure~\ref{fig:density}, we show the exact limiting distribution of the
highest bid for the binned LOB with uniform arrivals over 
50 bins, along with the limiting distribution for the continuous LOB. Note the ``shoulder'' bin: in the binned LOB, the threshold happens to fall into the middle of a bin, so the long-term probability of having the rightmost bid in the bin is positive but below the continuous limit. 

While we have been able to compute analytically the
distribution of the location of the rightmost bid, there are many related
quantities for which we do not have exact expressions in steady-state (although the
positive recurrence established 
in Theorem~\ref{thm: threshold} implies that they are
well-defined and can be estimated consistently from simulation). Notably,
except in the special case to be  considered in 
Section~\ref{sec:one-sided}, 
we have not been able to derive analytic expressions for the equilibrium
height of the book (i.e. expected number of bids or asks at a given price
in the binned model), or for the joint distribution of the highest bid and lowest ask. 
For an illustration of the 
simulated joint density of the highest bid and lowest ask,
see~\cite{EY_thesis}. 

\subsection{Brief summary of notation}\label{notation}
We summarize here our notation, as well as some of the main assumptions used in the text.

\noindent
$\CL$: limit order book.\\
$\CP$: price equivalence function, a monotone increasing function. Most of the time we use either $\CP(x) = x$ or the function that places all prices into one of several bins.\\
$A_t$, $B_t$: the counting measure of asks, respectively bids, at time $t$.\\
$F_a$, $F_b$: CDFs of the prices of arriving ask and bid orders. Until Section~\ref{sec:differing}, newly arriving orders are assumed to have equal probability of being a bid or an ask. In a binned model, we may write $F_{a,b}(n)$ (with $n$ an integer) to refer to the fraction of orders arriving into bins with index $\leq n$, i.e. the CDF evaluated at the rightmost endpoint of the interval.\\
$f_a$, $f_b$: the corresponding densities, which are assumed to exist. For
most results, $f_a$ and $f_b$ are assumed to be bounded above and below. \\
$\alpha_t$, $\beta_t$: the price of the lowest ask, respectively highest bid, at time $t$. Note that this is the actual price, not the bin containing it.\\
$\bin{x}$: in a binned LOB, the index of the bin containing price $x$.\\
$\kappa_a$, $\kappa_b$: limiting prices above (respectively below) which only finitely many asks (bids) are ever executed. It is not obvious a priori that $\kappa_a < 1$ or $\kappa_b > 0$; we prove this fact in Step 3 of the proof of Proposition~\ref{prop: weak distro}.\\
For functions of two or more arguments, we may interchange arguments and subscripts: thus, $f_{k,n}(t) \equiv f_n(k,t) \equiv f(k,n,t)$. We will use notation $f_n(k,\cdot)$ when we wish to consider $f$ as a function of the third argument alone.

\section{Preliminary results: monotonicity.}\label{sec:scaffold}
Before proving the main results, we erect some scaffolding. Part of its purpose is to allow us to transition between continuous LOBs (for which we expect to get differential equations in the answer) and binned models (which can be modeled as countable-state Markov chains). It will also allow us to compare LOBs with different arrival price distributions.

Lemma~\ref{lm: add one} asserts that the state of the limit order book is Lipschitz in the initial state with Lipschitz constant 1: in particular, small perturbations in the arrival and matching patterns will lead to small perturbations in the state of the book.
Lemma~\ref{lm: decrease queues} asserts that actions that decrease
cumulative bid and ask queues by either shifting orders or removing them in
bid--ask pairs will only decrease future queue sizes.

\begin{lemma}[Adding one order]\label{lm: add one}
Consider a limit order book $\CL$, and let $\tilde \CL$ differ from $\CL$ by the addition of one bid at time 0; let their arrival processes and price 
equivalence 
functions be the same. Then at all times $\tilde \CL$ differs from $\CL$ either by the addition of one bid or by the removal of one ask.
\end{lemma}
\proof{Proof.}
The roles of ``bid'' and ``ask'' are symmetric here. The claim clearly holds until the additional bid is the highest bid that departs from the system; once it does, $\CL$ differs from $\tilde \CL$ by the addition of a single ask, and the result follows by induction.\qedhere
\endproof

Define cumulative queue sizes $Q_b(p,t) = B_t(0,p]$, $Q_a(p,t) = A_t[p, 1)$. (Note that we count bids from the left and asks from the right.) When we want to highlight the dependence on only one of the variables, we will drop the other variable into a subscript.
\begin{lemma}[Decreasing queues]\label{lm: decrease queues}
Consider a limit order book $\CL$, and let $\tilde \CL$ differ from $\CL$ by modifying the initial state in such a way that $\tilde Q_b(\cdot,0) \leq Q_b(\cdot,0)$, $\tilde Q_a(\cdot,0) \leq Q_a(\cdot,0)$ (as functions of price), and also $\tilde Q_b(1,0) - \tilde Q_a(0,0) = Q_b(1,0) - Q_a(0,0)$. In words, to get from $\CL$ to $\tilde \CL$, at time 0 we remove some bid--ask pairs, and/or shift some bids to the right, and/or shift some asks to the left. Then at all future times $t \geq 0$, $\tilde Q_b(\cdot,t) \leq Q_b(\cdot,t)$ and $\tilde Q_a(\cdot,t) \leq Q_a(\cdot,t)$ as functions of price.
\end{lemma}

\proof{Proof.}
We show $\tilde Q_b \leq Q_b$, the argument for asks being identical. The argument proceeds by induction on time, i.e. the number of arrived orders. Throughout the proof, we use notation $f_{t-} = \lim_{s \searrow t} f(s)$ for the left limit of a c{\`a}dl{\`a}g function $f$.

Consider first the arrival of a bid at time $t$ and price $p$. For it to upset the inequality, it must stay in $\tilde\CL$ but depart immediately in $\CL$; additionally, we need $Q_b(q,t-) = \tilde Q_b(q,t-)$ for some $q \geq p$. Note that if the bid departs immediately in $\CL$, the leftmost ask at $\alpha_{t-}$ must be compatible with $p$, and in particular there are no bids right of $p$: $Q_b(p,t-) = Q_b(1,t-)$. This, together with $Q_b(q,t-) = \tilde Q_b(q,t-)$ and $\tilde Q_b(\cdot,t-) \leq Q_b(\cdot, t-)$, implies that $Q_b(1,t-) = \tilde Q_b(1,t-)$. Since bid--ask departures occur in pairs, this in turn implies $Q_a(0,t-) = \tilde Q_a(0,t-)$. But it is easy to see that if $\tilde Q_a(\cdot,t-) \leq Q_a(\cdot,t-)$ and they are equal at $0$, then $\tilde \alpha_{t-}$ (the leftmost jump of $\tilde Q_a(\cdot,t-)$) and $\alpha_{t-}$ (the leftmost jump of $Q_a(\cdot,t-)$) satisfy $\tilde \alpha_{t-} \leq \alpha_{t-}$, and hence the arriving bid actually departs immediately in $\tilde\CL$ as well.

Next consider the arrival of an ask at time $t$ and price $p$. For it to upset the inequality, it must cause the departure of the highest bid in $\CL$, but not in $\tilde\CL$, and we must have $Q_b(q,t-) = \tilde Q_b(q,t-)$ for some $q \geq \beta_{t-}$ with $\CP(\beta_{t-}) \geq \CP(p)$. Now, in $\tilde \CL$ there are no bids at prices $\geq \CP(p)$, hence $\tilde Q_b(1,t-) = \lim_{\epsilon \to 0}\tilde Q_b(p-\epsilon, t-)$. However, this contradicts the inequality $\tilde Q_b(\cdot, t-) \leq Q_b(\cdot,t-)$, since $\lim_{\epsilon \to 0} Q_b(p-\epsilon, t-) \leq \lim_{\epsilon \to 0} Q_b(\beta_{t-}-\epsilon, t-) \leq Q_b(1,t-) - 1$. (Note that inequalities may not be equalities if there are multiple bids at the same price.)\qedhere
\endproof

We can use this lemma to compare two limit order books $\CL$ and $\tilde \CL$ with identical initial states and order arrival processes, but different price equivalence functions. Suppose the price equivalence function $\tilde \CP$ merges some of the values that were distinguished by $\CP$. Then any bid--ask pair that is compatible in $\CL$ is also compatible in $\tilde \CL$; and possibly additional bid--ask pairs are compatible in $\tilde \CL$ as well. This lets us apply Lemma~\ref{lm: decrease queues} to conclude that fewer orders will be present in $\tilde \CL$.

We can give an upper bound on the queue sizes in $\CL$ by using a binned
LOB with one more bin and a shifted arrival process. If $\CL$ has a bid arrival at price $x$
in bin $k = \bin{x}$, we let $\widehat \CL$ have a bid arrival at some
price in bin $k-1$. The ask arrivals are identical in $\CL$ and $\widehat \CL$. (If bins of $\CL$ are numbered 1 through $N$, then bins of $\widehat \CL$ are numbered $0$ through $N$; bids arrive into bins 0 through $N-1$ in $\widehat \CL$, while asks arrive into bins 1 through $N$.) Under this arrangement, any bid--ask pair that is compatible in
$\widehat \CL$ was compatible in $\CL$ as well, so $\tilde \CL$ offers an
upper bound on the queue sizes of $\CL$. Consequently we can bound a continuous LOB $\CL$ both from above and from below by two binned LOBs with slightly different arrival price distributions. (Assuming the continuous LOB has arrival distributions supported on $[0,1]$, the binned LOB providing the upper bound will have bid arrival distribution supported on $[-\epsilon, 1-\epsilon]$ and ask arrival distribution supported on $[0,1]$.)

Finally, when bin sizes are small, the difference in the arrival price
distributions will be small, and we'll use Lemma~\ref{lm: add one} to bound
the rate at which the states of $\tilde \CL$ and $\widehat \CL$ diverge. This will let us show that the behavior of the continuous LOB converges in a suitable sense.

\section{Proof of main results.}\label{sec:proofs}

We begin by stating a weaker form of Theorem~\ref{thm: threshold}.
\begin{proposition}[Weak thresholds]\label{prop: weak phase}
There exist prices $\kappa_b$ and $\kappa_a$ with the following properties:
\begin{enumerate}
\item For any $\epsilon > 0$ there exists, almost surely, a (random) time $T_0 < \infty$ 
such that $\beta_t > \kappa_b$ and $\alpha_t < \kappa_a$ for all $t \geq T_0$.
\item For any $\epsilon > 0$, infinitely often there will be no bids with
price exceeding $\kappa_b + \epsilon$. Similarly, infinitely often
there will be no asks with price below $\kappa_a - \epsilon$.
\item The threshold values $\kappa_b$ and $\kappa_a$ satisfy $F_b(\kappa_b) = 1-F_a(\kappa_a)$.
\end{enumerate}
In addition, suppose that the bid and ask price densities (exist and) are bounded above by $M$. Then the following holds:
\begin{enumerate}
\setcounter{enumi}{3}
\item\label{part 4} For any $\epsilon > 0$, with probability 1, 
there exists a sequence of times $T_n \to \infty$ such that at time $T_n$ there are no bids with prices above $\CP(\kappa_b) + \epsilon$, and the number of asks with prices below $\CP(\kappa_a) - \epsilon$ is bounded above by $2(M+1)\epsilon T_n$.
\end{enumerate}
\end{proposition}

\begin{remark}
Although the compatibility of bid--ask pairs is driven by the price equivalence function $\CP(x)$, the statements about $\kappa$ are in terms of $x$ itself. This is because whenever there are compatible bid--ask pairs, the bid with the highest value of $x$ and the ask with the lowest value of $x$ always depart the book. In particular, in a binned model, $\kappa_b$ and $\kappa_a$ will usually fall in the middle of some bin; in this ``shoulder'' bin, a nontrivial fraction of the arriving orders remain in the book forever. In Figure~\ref{fig:density}, $\kappa_b$ is approximately half-way through the ``shoulder'' bin. (It is also possible for $\kappa_b$ to form the edge of a bin.)
\end{remark}

\begin{remark}
Note that we make no assertions here about the behavior of $n^{-1} T_n$ as $n \to \infty$: for the purposes of this proposition, this sequence may well tend to zero. The proof of Theorem~\ref{thm: threshold} will show that for any $\epsilon > 0$ the sequence $n^{-1} T_n = n^{-1} T_n(\epsilon)$ is in fact eventually bounded away from zero, with the bound depending on $\epsilon$.
\end{remark}

\proof{Proof.}
The first two claims follow from Kolmogorov's 0--1 law. Consider the events
\begin{align*}
&\CE_b(x) = \{\text{finitely many bids will depart from prices $\leq x$}\}\\
&\CE_a(x) = \{\text{finitely many asks will depart from prices $\geq x$}\}. 
\end{align*}
Lemma~\ref{lm: add one} shows that these events are in the tail $\sigma$-algebra of the arrival process. Since the arrival process consists of 
a sequence of independent and identically distributed events, 
Kolmogorov's 0--1 law ensures that for each $x$, $\CE_b(x)$ has probability 0 or 1 (and similarly for $\CE_a(x)$). Now let
\begin{subequations}\label{eqns: kappa}
\begin{equation}\label{eqn: 0--1 kappa}
\kappa_b = \sup\{x: \BP(\CE_b(x)) = 1\}, \qquad \kappa_a = \inf\{x:
\BP(\CE_a(x)) = 1\}.
\end{equation}
(If the set whose extremum is to be taken is empty, we let $\kappa_b = 0$ or $\kappa_a = 1$.)
The first two asserted properties now follow upon noticing that $\CE_b(x) \subseteq \CE_b(y)$ for $x \geq y$, and that whenever there is a bid departure at price $x$, there must be no bids at prices higher than $x$. (The situation is similar for asks.)


We next show that $F_b(\kappa_b) + F_a(\kappa_a) = 1$. From the strong 
law of large numbers for the arrival process and the 0--1 law above, we know that $F_b(\kappa_b)$ is the smallest limiting proportion of arriving bids that stay in the system:
\begin{equation}\label{eqn: LLN kappa}
F_b(\kappa_b) = \lim\inf_{t \to \infty} \frac1t \#(\text{bids in the LOB at time $t$}).
\end{equation}
\end{subequations}
A similar equality clearly holds for asks with $1-F_a(\kappa_a)$. Since
bids and asks always depart in pairs, a further appeal to 
the strong law of large numbers for the arrival process shows that 
we must have $F_b(\kappa_b) = 1-F_a(\kappa_a)$.

The existence of times $T_n$ as in part (4) of the theorem 
follows by a similar argument from the functional law of large numbers for the arrival processes. With probability 1, picking a large enough time $T_n$ when there are no bids at prices above $\CP(\kappa_b) + \epsilon$ ensures that there are at most $(F_b(\kappa_a) + (M+1)\epsilon)T_n$ asks in the system. Since asks to the right of $\kappa_a$ arrive at rate $(1-F_a(\kappa_a)) = F_b(\kappa_b)$ and eventually never leave, for large enough $T_n$ there will be at most $2(M+1)\epsilon T_n$ asks at prices below $\kappa_a - \epsilon$.\qedhere
\endproof

This result is weaker than the positive recurrence we wish to prove
eventually: in particular, it does not 
show that the total number of both types of orders between $\kappa_b$ and $\kappa_a$ is ever zero. 
To obtain
statements about positive recurrence, we shall need to use fluid limit
techniques, and our overall approach will be similar to that 
in \cite[Chapter 4]{Bramson}. The final proof of stability
will use multiplicative Foster's criterion
(state-dependent drift), see \cite[Theorem 13.0.1]{Meyn_Tweedie}. In order to
get there, we need to show that whenever there are many bids or asks in
$(\kappa_b, \kappa_a)$, their number tends to decrease at some positive, bounded
below,
 rate over long periods of time. This is a standard line of argument in
queueing theory; but the challenge of the model is that the evolution of the queues depends on which queues are positive, rather than which queues are large. In general Markov chains of this form are very difficult to analyze (\cite{Gamarnik_Katz} show that in general the stability of such chains is undecidable), but the special structure of our chain makes it amenable to analysis. The outline of the proof is as follows.
\begin{enumerate}
\item We work with binned LOBs. We begin by showing that, after appropriate
rescaling, both the queue sizes and the local time of the highest bid
(lowest ask) in each bin converge to a set of Lipschitz trajectories, which
we call \emph{fluid limits}. We then proceed to develop 
properties of the fluid limits.
\item We next show that all fluid limits tend to zero for bins (strictly) between $\bin{\kappa_b}+1$ and $\bin{\kappa_a}-1$. We exploit the equations and inequalities satisfied by fluid limits to show the following:
\begin{enumerate}
\item There is an interval $[x_0,y_0]$ on which, whenever the fluid limit of the number of orders is positive, it decreases (at a rate bounded below). Therefore, after some time $T_0$ (which depends on the initial state), the fluid limit will be zero on $[x_0,y_0]$. The values $x_0$ and $y_0$ may not be bin boundaries.
\item Following $T_0$, we will be able to bound from below the rate of increase of the local time of the rightmost bid on $[x_1,x_0)$ for some $x_1 < x_0$, and of the leftmost ask on $(y_0,y_1]$ for some $y_1 > y_0$. Since whenever the highest bid is in $[x_1,x_0)$ it has a positive chance of departing (and similarly for asks in $(y_0,y_1]$), we conclude that whenever the number of orders in $[x_1,y_1]$ is large, it will decrease (at a rate bounded below). We repeat the argument until $[x_n,y_n] \approx [\kappa_b,\kappa_a]$. The $x_i$ and $y_i$ may not be bin boundaries in this step.
\end{enumerate}
\item We show that if on some interval, all fluid limits converge to 0 in finite time, then the binned LOB is recurrent on that interval. (This step is standard for fluid limit arguments.) Since the number of bids in a continuous limit order book can be bounded from above by binned ones, this will also show recurrence of the continuous LOB.
\end{enumerate}

\subsection{ODE of the limiting distribution.}
Our first result shows that the ODE which should describe the unique limit,
as $t \to \infty$, of the empirical distribution of the highest bid does in
fact describe \emph{some} such limit. In the process, we also establish $0
< \kappa_b < \kappa_a < 1$. 

\begin{proposition}[Weak distribution of the highest bid]\label{prop: weak distro}
Suppose the arrival price distributions have densities bounded above and
below, and consider a sequence of binned LOBs with the number of bins, $N$, tending to infinity.

For each $N$ and $\epsilon > 0$, let $T_n = T_n(N,\epsilon) \to \infty$ be the sequence of times identified in part \eqref{part 4} of Proposition~\ref{prop: weak phase}. Let $\pi_b(n,N,\epsilon)$ be the discrete normalized empirical density of the highest bid over the time interval $[0,T_n]$; that is,
\[
\pi_b(n,N,\epsilon,x) = \frac{\text{time up to $T_n$ that the highest bid is in $\bin{x}$}}{T_n \cdot \text{(length of $\bin{x}$)}}.
\]
\begin{enumerate}
\item There exists a unique limit $\lim_{n \to \infty,\ N \to \infty,\ \epsilon \to 0} \pi_b(n,N,\epsilon) := \pi_b$, and similarly for asks.
\item Denoting $\varpi_b = \pi_b/f_b$ and $\varpi_a = \pi_a/f_a$, these satisfy the pair of integral equations
\begin{subequations}\label{eq:integral}
\begin{equation}
F_a(x)\varpi_b(x) = \int_x^1 \varpi_a(y) f_a(y) dy,~~ x \in (\kappa_b, \kappa_a); \qquad \int_{\kappa_b}^{\kappa_a} \varpi_b(x) f_b(x) dx = 1,
\end{equation}
\begin{equation}
(1-F_b(x))\varpi_a(x) = \int_0^x \varpi_b(y) f_b(y) dy,~~ x \in (\kappa_b,\kappa_a); \quad \int_{\kappa_b}^{\kappa_a} \varpi_a(x) f_a(x) dx = 1.
\end{equation}
\end{subequations}
\item Moreover, wherever $\varpi_b$ is differentiable, it satisfies the ODE
\begin{subequations}\label{eq: ODE}
\begin{equation}\label{eq: diffeq}
\left(-\frac{1-F_b(x)}{f_a(x)}(F_a(x)\varpi_b(x))'\right)' = \varpi_b(x)f_b(x)
\end{equation}
with initial conditions
\begin{equation}\label{eq: initial}
(F_a(x)\varpi_b(x)) \vert_{x = \kappa_b} = 1, \qquad (F_a(x)\varpi_b(x))'\vert_{x=\kappa_b} = 0
\end{equation}
\end{subequations}
and the additional constraint $\varpi_b(x) \to 0$ as $x \uparrow \kappa_a$. The distribution of the leftmost ask satisfies a similar ODE.
\item The equation \eqref{eq: ODE} has a unique solution; in particular, $\kappa_b$ and $\kappa_a$ are uniquely determined by it.
\end{enumerate}
\end{proposition}

\begin{remark}[Normalization and initial conditions]\label{rem: density 1/x}
From the integral equation \eqref{eq:integral} it follows that $\varpi_b$ will be a continuous function of price, whereas $\pi_b$ may not be. In particular, if we are interested in piecewise continuous functions $f_b$ and $f_a$, then $\varpi_b$ will satisfy the ODE \eqref{eq: diffeq} on each of the segments where $f_b$ and $f_a$ are continuous, and can be patched together from the requirement that $\varpi_b(x)$ and $(F_a(x)\varpi_b(x))'$ are both continuous.

The initial conditions \eqref{eq: initial} apply for LOBs with finite
initial states. Consider instead a LOB $\tilde \CL$ with an infinite bid
order at some price $p > \kappa_b$. As long as the threshold $\tilde
\kappa_b$ of $\tilde \CL$ is positive, we can do away with the infinite
order at $p$ by changing the price equivalence function so that $\CP(0) =
\CP(\tilde\kappa_b) =  \CP(p)$: the evolution of $\tilde\CL$ and this new
LOB $\hat\CL$ will be the same at prices above $p$ after the threshold
time. In $\hat\CL$, there is yet another threshold $\hat\kappa_b$, and the
initial conditions \eqref{eq: initial} hold for all $x \in (\hat\kappa_b,
p]$, meaning $\hat\varpi_b(x) = 1/F_a(x)$ on that interval.
Correspondingly, in $\tilde \CL$, the distribution of the highest bid price
has an atom at $p$ of mass $\int_{\hat \kappa_b}^p 1/F_a(x) dx$. For the lowest ask price, we will of course have $\BP(\alpha_t \leq p) = 0$, but it may be the case that $\varpi_a(x)\not\to 0$ as $x \downarrow p$: it may be discontinous at the location of the infinite bid.
\end{remark}

\begin{remark}
The computations in our Steps~\ref{ass2} and \ref{ass2.5} below are 
similar to
computations appearing in \cite[Section 1]{Luckock}; we present the
full argument for completeness.
\end{remark}

\proof{Proof.}
The proof proceeds as follows:
\begin{enumerate}
\item\label{ass1} Fix the number of bins $N$, and consider the collection of empirical densities $\pi_b(n,N,\epsilon)$, $\pi_a(n,N,\epsilon)$. Along any sequence $n \to \infty$, $N \to \infty$, and $\epsilon \to 0$ there is a convergent subsequence.
\item\label{ass2} Any subsequential limit satisfies a certain pair of integral equations, hence some ODEs.
\item\label{ass2.5} The ODEs will directly imply $\kappa_b < \kappa_a$; in addition, $0 < \kappa_b$ and $\kappa_a < 1$.
\item\label{ass3} The solution to these ODEs is unique, and in particular the limit does not depend on the order of $n \to \infty$, $N \to \infty$, and $\epsilon \to 0$.
\end{enumerate}



{\bf Step 1:} The space of probability distributions with compact support is compact, so along any sequence of empirical distributions there will be convergent subsequences. Moreover, whenever the highest bid is in bin $\bin{x}$, bid departures occur from the bin at rate $\geq F_a(x) \pi_b(\bin{x})$, whereas bid arrivals occur into that bin at rate at most $f_b(x)$. Consequently, under the assumption of bounded densities $f_b, f_a$, the highest bid density $\pi_b(\bin{x}) \leq f_b(x) / F_a(x)$ is bounded uniformly in $n$, $N$, $\epsilon$; this guarantees the existence of limiting \emph{densities} along subsequences. Finally, the lower bound on $f_a$ and $f_b$ guarantees that $\varpi_a$ and $\varpi_b$ are bounded, and hence also converge along subsequences. For steps 2 and 3, $\pi_{a,b}$ and $\varpi_{a,b}$ refer to any such subsequential limit, taken along a single subsequence for all four quantities.

{\bf Step 2:} The integral equations are expressing the idea that the rate of bid arrival should be equal to the rate of bid departure. Along a sequence of times where the queues are small (i.e. $\epsilon \approx 0$), this is very nearly true; it will be exactly true in the limit $\epsilon \to 0$. The bid arrival rate at $x$ is $f_b(x) \BP(\alpha_t > x) = f_b(x) \int_x^1 \pi_a(y)dy$, and the bid departure rate at $x$ is $\pi_b(x) F_a(x)$, so setting the two equal gives the result; the ODE is obtained by differentiating twice.

Of course, if we fix the number of bins $N$, the limit distribution will be described by a difference equation rather than an integral (or differential) equation. It is standard to see that the limit of solutions to the difference equations solves the differential (or integral) equation.

{\bf Step 3:} To see $\kappa_b < \kappa_a$, note that $\pi_b$ is bounded above by $f_b/F_a$ always, so if it integrates to 1 we must have $\kappa_b < \kappa_a$. To see $\kappa_b > 0$ (and $\kappa_a < 1$), we consider a binned LOB $\tilde\CL$ with three bins, with bin partitions at $x$ and $x + \delta$ for some $x \in (\kappa_b,\kappa_a)$. By monotonicity, $\bin{\tilde\kappa_b} = 1$ and $\bin{\tilde\kappa_a} = 3$. For $\delta$ small enough, the number of orders in the middle bin will eventually be stochastically dominated by a geometric random variable. Indeed, whenever there are bids in bin 2, more bids arrive at rate $F_b(x+\delta) - F_b(x)$ and depart at the larger rate $F_a(x+\delta)$ (this is after asks from bin 3 stop departing). The situation is similar for asks. Consequently, in $\tilde\CL$ we must have $\tilde\pi_b(2) > 0$ and $\tilde\pi_a(2) > 0$.

If $\tilde\pi_b(1)$ and $\tilde\pi_a(3)$ were such that (almost) all orders depart, then from $\tilde\pi_b(1) F_a(x) = F_b(x)$ we find
\[
\tilde\pi_b(1) F_a(x) = F_b(x) \implies \tilde\pi_b(2) = \frac{F_a(x) - F_b(x)}{F_a(x)} \implies F_a(x) > F_b(x).
\]
Now let $\delta$ be small enough that $F_a(x) > F_b(x+\delta)$, and solve for $\tilde\pi_b(2)$ from the alternative expression $\tilde\pi_b(2)F_a(x+\delta) = (F_b(x+\delta) - F_b(x))\tilde\pi_a(3)$. This gives
\[
\tilde\pi_b(1) + \tilde\pi_b(2) = \frac{F_b(x)}{F_a(x)} + \frac{F_b(x+\delta)-F_b(x)}{F_a(x+\delta)}\frac{1-F_a(x+\delta)}{1-F_b(x+\delta)} <1.
\]
The contradiction shows that in fact in this LOB we must have $\tilde\pi_b(1) F_a(x) < F_b(x) - \eta$ for some $\eta > 0$, which implies $F_b(\tilde\kappa_b) \geq \eta$. By monotonicity, we obtain $\kappa_b > 0$ as well (for $N$ large enough that the above bin of width $\delta$ is one of the original bins of the LOB).

{\bf Step 4:}
The uniqueness of solution follows from the fact that we have a second-order ODE with two initial conditions (which, as we just showed, are finite). Note that an alternative argument for $\kappa_b > 0$ would be to show that $\pi_b(x) > 0$ for some $x > 0$, since then the ODE forces $\pi_bF_a/f_b$ decreasing, and $\pi_b(x) \sim 1/x$ near 0, which is not integrable. However, it is not immediately obvious why in a binned LOB the highest bid couldn't spend (almost) all of its time in the leftmost bin, hence we give the more involved argument above.

It is at this moment possible that there are multiple solutions to the ODE with different values for $\kappa_b$. Intuitively, this should not be the case, since any limiting $\kappa_b$ should give the (unique) threshold value of the continuous LOB. We will derive the uniqueness of the quadruple $(\kappa_b, \kappa_a, \pi_b, \pi_a)$ from Lemma~\ref{lm: ODE monotonicity} below, which shows that the solution of \eqref{eq: diffeq} is monotonic in the initial conditions. This implies that the requirements $\int_{\kappa_b}^{\kappa_a}\pi_b(x)dx = 1$ and $F_b(\kappa_b) = 1-F_a(\kappa_a)$ pin down $\kappa_b$ and $\kappa_a$ uniquely, since decreasing $\kappa_b$ increases the initial value of $\varpi_b$ and $\frac{d}{dx}\varpi_b$.\qedhere
\endproof

The second result we require about the ODE is monotonicity in the initial conditions:
\begin{lemma}[ODE monotonicity]\label{lm: ODE monotonicity}
Let $\varpi_b$ and $\tilde\varpi_b$ be two solutions of the ODE \eqref{eq: diffeq} with initial conditions
\[
\varpi_b(x_0) \geq \tilde\varpi_b(x_0), \qquad (\varpi_b)'(x_0) \geq (\tilde\varpi_b)'(x_0).
\]
Then for all $x \geq x_0$, $\varpi_b(x) \geq \tilde\varpi_b(x)$.
\end{lemma}
\proof{Proof.}
We may reparametrize space monotonically so that $F_a(x) = x$. Then the ODE \eqref{eq: diffeq} becomes
\[
(F_b(x)-1)\left(x \frac{d}{dx}\varpi_b(x) + \varpi_b(x)\right)' + x f_b(x) \frac{d}{dx}\varpi_b(x) = 0,
\]
which is a first-order ODE in $\frac{d}{dx}\varpi_b(x)$. Since solutions of first-order ODEs are increasing in their initial conditions, we obtain $\frac{d}{dx}\varpi_b(x) \geq \frac{d}{dx}\tilde\varpi_b(x)$, and the desired inequality follows trivially.\qedhere
\endproof
\begin{corollary}\label{coroll: ODE monotonicity}
Suppose the initial conditions for $\varpi_b$ come \eqref{eq: initial}, and the initial conditions for $\tilde\varpi_b$ are $F_a(x_0)\tilde\varpi_b(x_0) = 1$, $(F_a(x)\tilde\varpi_b(x))' \vert_{x=x_0} = 0$ for some $x_0 > \kappa_b$. Then $\varpi_b(x_0) \leq \tilde\varpi_b(x_0)$ and $\frac{d}{dx}\varpi_b(x)\vert_{x=x_0} \leq \frac{d}{dx}\tilde\varpi_b(x)\vert_{x=x_0}$. Consequently, $\varpi_b(x) \leq \tilde\varpi_b(x)$ for $x \geq x_0$.
\end{corollary}
\proof{Proof.}
Reparametrize space as before, so that $F_a(x) = x$. Then
\[
(x\varpi_b(x))' = -\pi_a(x), \quad \varpi_b(x) = \frac1x \left(1 - \int_0^x \varpi_a(y)dy \right).
\]
From this it is clear that $\varpi_b(x_0) \leq \tilde\varpi_b(x_0)$. Further,
\[
x \frac{d}{dx}\varpi_b(x) = -\pi_a(x) - \varpi_b(x) = - \frac1x +\frac1x \int_0^x(\varpi_a(y) - \varpi_a(x)) dy.
\]
Now, in a LOB, $(1-F_b(x))\varpi_a(x)$ is increasing (cf. $x \varpi_b(x)$ which is decreasing), meaning $\varpi_a$ is increasing. Consequently, the integral above is nonpositive, and we see
\[
x_0 \frac{d}{dx}\varpi_b(x)\vert_{x=x_0} \leq -\frac{1}{x_0} = x_0 \frac{d}{dx}\tilde\varpi_b(x)\vert_{x=x_0}
\]
as required.\qedhere
\endproof

\subsection{Fluid limits.}
In this section we introduce the fluid-scaled processes associated 
with the limit order book, discuss their convergence to fluid limits, and determine properties of the limits. Throughout the section, we work with a binned limit order book.

Let $B_k(\cdot)$ and $A_k(\cdot)$ be the arrival processes of bids and asks into bin $k$ (indexed by time). The time structure of these processes is not important for our results, so we may assume that these are Poisson processes; by definition, they are independent. We will assume that the total arrival rate of bids is 1, and also of asks, so that if $p_b(k)$ (respectively $p_a(k)$) is the probability that an arriving bid (ask) falls into bin $k$, this is also the arrival rate of bids (asks) into that bin. Let $Q_b(k,t)$ (respectively $Q_a(k,t)$) be the number of bids (asks) in bin $k$ at time $t$. Let $T_\beta(k,t)$ and $T_\alpha(k,t)$ be the amount of time up to time $t$ when the rightmost bid, respectively leftmost ask, is in bin $k$: that is,
\[
T_\beta(k,t) = \int_0^t \one\{\bin{\beta_s} = k\} ds, \qquad T_\alpha(k,t) = \int_0^t \one\{\bin{\alpha_s} = k\} ds.
\]
It is clear that the initial data $Q_b(k,0)$, $Q_a(k,0)$ together with the arrival processes $B_k(\cdot)$, $A_k(\cdot)$ give sufficient information to determine the values of all of these processes at later times. We have the following expressions:
\begin{subequations}\label{eq:integral1}
\begin{align}
&\bin{\beta_t} = k \iff Q_b(k,t) > 0, \quad \sum_{k' > k}Q_b(k',t) = 0\label{helper1}\\
&\bin{\alpha_t} = k \iff Q_a(k,t) > 0, \quad \sum_{k' < k}Q_a(k',t) = 0\label{helper2}\\
&Q_b(k,t) = Q_b(k,0) + \int_0^t \one\{\bin{\alpha(s)} > k\} dB_k(s) - \sum_{k' \leq k} \int_0^t \one\{\bin{\beta(s)} = k\} dA_{k'}(s)\\
&Q_a(k,t) = Q_a(k,0) + \int_0^t \one\{\bin{\beta(s)} < k\} dA_k(s) - \sum_{k' \geq k} \int_0^t \one\{\bin{\alpha(s)} = k\} dB_{k'}(s)\label{int4}\\
&T_\beta(k,t) = T_\beta(k,0) + \int_0^t \one\{\bin{\beta(s)} = k\} ds\\
&T_\alpha(k,t) = T_\alpha(k,0) + \int_0^t \one\{\bin{\alpha(s)} = k\} ds
\end{align}
\end{subequations}

We define the \emph{fluid-scaled} processes by $\ov X_n(t) = n^{-1}X(nt)$ for any process $X$. We now have the following result on convergence to fluid limits:
\begin{theorem}[Convergence to fluid limits]\label{thm: fluid convergence}
Consider a sequence of processes
\[
(\ov B_n(k,\cdot), \ov A_n(k,\cdot), \ov Q_{b,n}(k,\cdot), \ov Q_{a,n}(k,\cdot), \ov T_{\beta,n}(k,\cdot), \ov T_{\alpha,n}(k,\cdot))
\]
whose initial state (at time 0) is bounded: $||\ov Q_{a,n}(k,0), \ov Q_{b,n}(k,0)|| \leq 1$. As $n \to \infty$, any such sequence has a subsequence which converges, uniformly on compact sets of $t$, to a collection of Lipschitz functions
\[
(b_k(\cdot), a_k(\cdot), q_b(k,\cdot), q_a(k,\cdot), \tau_\beta(k,\cdot), \tau_\alpha(k,\cdot)).
\]
(Different subsequences may converge to different 6-tuples of Lipschitz functions.) We call the limiting 6-tuple a fluid limit.

Any fluid limit satisfies the following equations almost everywhere (i.e. everywhere where the derivatives are defined):
\begin{subequations}\label{eq: limit eqs}
\begin{align}
&b'_k(t) = p_b(k), \quad a'_k(t) = p_a(k)\label{eq1}\\
&\frac{\del}{\del t}\left(\tau_\beta(k,t)\right) = 0 \text{ if } \sum_{k' > k}q_b(k',t) > 0, \quad \frac{\del}{\del t}\left(\tau_\alpha(k,t)\right) = 0 \text{ if } \sum_{k' < k}q_a(k',t) > 0\label{eq2}\\
&\sum_{k = \bin{\kappa_b}-1}^{\bin{\kappa_a}} \tau_\beta(k,t) = t, \quad \sum_{k = \bin{\kappa_b}}^{\bin{\kappa_a}+1} \tau_\alpha(k,t) = t\label{eq3}\\
&q_b(k,t) \geq 0, \quad q_a(k,t) \geq 0\label{eq4}\\
&\frac{\del}{\del t}q_b(k,t) = 0 \text{ if } q_b(k,t) = 0, \quad \frac{\del}{\del t}q_a(k,t) = 0 \text{ if } q_a(k,t) = 0\label{eq5}\\
&\frac{\del}{\del t}q_b(k,t) = p_b(k) \sum_{k' > k} \frac{\del}{\del t}\tau_\alpha(k',t) - \frac{\del}{\del t}\tau_\beta(k, t)\sum_{k' \leq k} p_a(k')\label{eq6a}\\
&\frac{\del}{\del t}q_a(k,t) = p_a(k) \sum_{k' < k} \frac{\del}{\del t}\tau_\beta(k',t) - \frac{\del}{\del t}\tau_\alpha(k, t)\sum_{k' \geq k} p_b(k').\label{eq6b}
\end{align}
\end{subequations}
\end{theorem}

\proof{Proof.}
The expressions in \eqref{eq:integral1} together with the functional law of large numbers for the arrival processes lead to the u.o.c. convergence along subsequences to a fluid limit. The integral representation implies that limits must be Lipschitz functions.

To see that any fluid limit must satisfy \eqref{eq: limit eqs}, we note that \eqref{eq1} follows directly for the functional law of large numbers for the arrival processes. Identities \eqref{eq2} follows from the corresponding statement for prelimit processes: if $\sum_{k' > k}q_b(k',s) > \epsilon > 0$ on a time interval $s \in (t-\epsilon,t+\epsilon)$, then for all sufficiently large $n$, $\sum_{k' > k} Q_{b, n}(k',ns) > n\epsilon/2 > 0$, so $\bin{\beta(ns)} > k$ and $T_{\beta, n}(k,ns)$ is not increasing. Identity \eqref{eq3} holds because the rightmost bid (leftmost ask) is eventually always in one of the bins in the prelimit processes, so this must be true in the limit. Identities \eqref{eq4} follows for a similar reason: prelimit queues are nonnegative, hence the limit is nonnegative as well.

Identity \eqref{eq5} is a corollary of \eqref{eq4}: a process that is always nonnegative, differentiable at $t$, and equal to 0 at $t$ must have derivative 0 there.

Finally, identities \eqref{eq6a} and \eqref{eq6b} follow from \eqref{helper1}--\eqref{int4} for the prelimit queues. More precisely, the rate at which the bid queue size changes is as follows: if the lowest ask is higher than bin $k$, then bids arrive into the queue at rate $p_b(k)$; and if the highest bid is in bin $k$, then all asks arriving at prices below $k$ deplete the queue at $k$. Because the location of the highest bid or lowest ask does not show up in the fluid limit, we instead use the local times $t_\beta$ and $t_\alpha$.\qedhere
\endproof

We introduce notation $\pi_\beta(k,t) = \frac{\del}{\del t}\tau_\beta(k,t)$, $\pi_\alpha(k,t) = \frac{\del}{\del t}\tau_\alpha(k,t)$.

\subsection{Fluid limits drain.}
We will now show that in a LOB that starts with infinitely many bids in $\bin{\kappa_b}+1$ and asks in $\bin{\kappa_a}-1$, the fluid limit queue sizes drain, i.e. converge to 0 on the bins ranging from $\bin{\kappa_b}+1$ to $\bin{\kappa_a}-1$. We will assume that bin widths (and hence $p_b(k)$, $p_a(k)$) are all small. This is the meat of the argument in the paper.

\begin{theorem}[Fluid limits drain]\label{thm: fluid stable}
Consider a fluid limit corresponding to a binned LOB with $N$ bins. Suppose the arrival process is symmetric ($p_b(k) = p_a(N-k)$), the probabilities $p_{a,b}(k)$ are bounded below, and $p_b(k)$ is decreasing in $k$ ($p_a(k)$ increasing in $k$). Suppose that initially there are infinitely many bids in bin $\bin{\kappa_b}+1$ 
and infinitely many asks in $\bin{\kappa_a}-1$; then the fluid limit of queues can be described by $q_{a,b}(k,t)$ for $\bin{\kappa_b}+2 \leq k \leq \bin{\kappa_a}-2$, and the fluid limit of the local times can be described by $\pi_{a,b}(k,t)$ for $\bin{\kappa_b}+1 \leq k \leq \bin{\kappa_a}-1$.

Let the initial state of the fluid limit satisfy $\norm{(q_b(0), q_a(0)} \leq 1$. There exists $\epsilon = \epsilon(N) \to 0$ as $N \to \infty$, and a time $T$ depending on $\{p_a(k), p_b(k), \text{bin widths}\}$, such that for all bins $k$ satisfying $\bin{\kappa_b + \epsilon} < k < \bin{\kappa_a - \epsilon}$, and all times $t \geq T$,
\[
q_b(k,t) = 0, \qquad q_a(k,t) = 0, \qquad \forall t \geq T.
\]

Further, in the interval $\bin{\kappa_b + \epsilon} < k < \bin{\kappa_a - \epsilon}$ and for $t \geq T$, the derivatives $\pi_\beta(k,t)$ satisfy the second-order difference equation
\[
\Delta_k\left(\frac{1-F_b(k)}{p_a(k+1)} \cdot \Delta_k\left(\frac{F_a(k)}{p_b(k)}\pi_\beta(k)\right)\right) = \pi_\beta(k+1),
\]
where the operator $\Delta_k$ is given by $\Delta_k(f) = f(k+1) - f(k)$. The initial conditions satisfy
\[
\frac{F_a(\bin{\kappa_b+\epsilon})}{p_b(\bin{\kappa_b+\epsilon})} \pi_\beta(\bin{\kappa_b+\epsilon}) \leq 1, \qquad \Delta_{\bin{\kappa_b+\epsilon}}\left(\frac{F_a(k)}{p_b(k)}\pi_\beta(k)\right) \leq 0. \notag
\]
A similar equation holds for asks. As $N \to \infty$, the solution of the difference equation converges to the solution of the ODE \eqref{eq: diffeq} with initial conditions given by \eqref{eq: initial}.
\end{theorem}
Note that $\kappa_a$ and $\kappa_b$ are the thresholds of an LOB with a finite starting state; the LOB with infinite bid and ask orders can be thought of as having different thresholds $\tilde \kappa_b < \kappa_b$ and $\tilde \kappa_a > \kappa_a$. For large $N$, Lemma~\ref{lm: add one} implies $\tilde\kappa_b \approx \kappa_b$ and $\tilde\kappa_a \approx \kappa_a$.

\proof{Proof.}
The proof proceeds in stages.

{\bf Stage 0.} Let $x_0$ be given by $F_a(x_0) = F_b(\kappa_a) - F_b(x_0)$, and let $y_0$ be given by $1-F_b(x_0) = (1-F_a(\kappa_b)) - (1-F_a(y_0))$. Equivalently, $F_a(x_0) + F_b(x_0) = 2x_0 = F_b(\kappa_a)$, so $x_0 = \frac12 F_b(\kappa_a)$, and $y_0 = \frac12(1+F_a(\kappa_b))$.

{\bf Claim 0.1:} $\kappa_b \leq x_0 < y_0 \leq \kappa_a$.\\
\emph{Proof:} Note that $F_a(\kappa_b)$ is a lower bound on the rate of bid
departure from the Markov chain when there are any bids present, while
$F_b(\kappa_b) - F_b(\kappa_a)$ is an upper bound on the rate of bid
arrival. Consequently, if $F_a(\kappa_b) > F_b(\kappa_b) - F_b(\kappa_a)$,
then the number of bids on the entire interval $(\kappa_b, \kappa_a)$ would
be stochastically bounded, whereas it should scale as a random walk. A
similar argument gives $y_0 \leq \kappa_a$. Finally, $\frac12 F_b(\kappa_a)
= \frac12(1-F_a(\kappa_b)) < \frac12(1+F_a(\kappa_b))$, since $\kappa_b >
0$ by Proposition~\ref{prop: weak distro}.\qedhere

{\bf Claim 0.2:} There exists $T_0 = T_0(M)$ such that for all times $t \geq T_0$ and all fluid models, $\sum_{k=\bin{x_0}+1}^{\bin{y_0}-1}(q_b(k,t) + q_a(k,t)) = 0$.\\
\emph{Proof:} Since these processes are absolutely continuous and nonnegative, it suffices to show that whenever there are any fluid orders in the interval (and all the derivatives are defined), the fluid number of orders in the interval decreases at a rate bounded below. By \eqref{eq6a} and \eqref{eq6b}, we see that for $Q(t) = \sum_{\bin{x_0}+1 \leq k \leq \bin{\kappa_a}-1} q_b(k,t)$,
\[
Q'(t) \leq \begin{cases}
0, & Q(t) = 0\\
\sum_{k=\bin{x_0}+1}^{\bin{\kappa_b}-1} p_b(k) - \sum_{k' \leq \bin{x_0}+1} p_a(k) < F_b(\kappa_b) - F_b(x_0) - F_a(x_0) - \epsilon, & Q(t) > 0.
\end{cases}
\]
Consequently, after a finite amount of time $T_{b,0}$, there will be no fluid bids in bins $\geq \bin{x_0}+1$. Similarly, after a finite amount of time $T_{a,0}$, there will be no fluid asks in bins $\leq \bin{y_0}-1$; we may take $T_0 = \max(T_{b,0}, T_{a,0})$.\qedhere

{\bf Claim 0.3:} There exists $\epsilon_0 > 0$ such that for all times $t \geq T_0$ and all fluid models, $\sum_{k \leq \bin{x_0}}\pi_\beta(k,t) \geq \epsilon_0$ and $\sum_{k \geq \bin{y_0}} \pi_\alpha(k,t) \geq \epsilon_0$. (This result requires bins to be sufficiently small.)\\
\emph{Proof:} Note that equations \eqref{eq6a} and \eqref{eq6b} hold at all times, even when there are no fluid orders in the bin; thus, for $t \geq T_0$ and all $k \in [\bin{x_0}+1,\bin{y_0}-1]$ we have
\[
p_b(k) \sum_{k' > k} \pi_\alpha(k',t) = \pi_\beta(k,t)\sum_{k' \leq k} p_a(k'), \qquad p_a(k) \sum_{k' < k} \pi_\beta(k',t) =  \pi_\alpha(k,t)\sum_{k' \geq k} p_b(k').
\]
Omitting the dependence on $t$ for clarity, these equations, together with the observation that $\sum_k \pi_\alpha(k) = \sum_k\pi_\beta(k) = 1$, can be rearranged to give two decoupled second-order difference equations for $\pi_\alpha(k)$ and $\pi_\beta(k)$. We abuse notation to write $F_a(k) = \sum_{k' \leq k} p_a(k')$ and similarly for $F_b(k)$.
\begin{subequations}\label{eqn:diffeq}
\begin{equation}\label{eqn:diff}
\Delta_k\left(\frac{1-F_b(k)}{p_a(k+1)} \cdot \Delta_k\left(\frac{F_a(k)}{p_b(k)}\pi_\beta(k)\right)\right) = \pi_\beta(k+1), \qquad \bin{x_0}+1 \leq k \leq \bin{y_0}-1.
\end{equation}
(There is a corresponding equation for $\pi_a$, of course.)

If we had two initial conditions for this second-order difference equation, we would be able to solve it. Unfortunately, in general we do not have such initial conditions, but we have bounds on them, namely
\begin{equation}\label{eqn:initial}
\frac{F_a(\bin{x_0})}{p_b(\bin{x_0})} \pi_\beta(\bin{x_0}) \leq 1, \qquad \Delta_{\bin{x_0}}\left(\frac{F_a(k)}{p_b(k)}\pi_\beta(k)\right) \leq 0.
\end{equation}
\end{subequations}
These inequalities would hold with equality in a different limit order book $\tilde\CL_0$, in which we assign the same low price to all the bins up through $\bin{x_0}+1$, and the same high price to all the bins from $\bin{y_0}-1$ up. (We nonetheless keep track the bins containing the highest bid and lowest ask of $\tilde\CL$.) Corollary~\ref{coroll: ODE monotonicity} shows that the solutions to \eqref{eqn:diffeq} on $\bin{x_0}+1 \leq k \leq \bin{y_0}-1$ are bounded from above by the solution for $\tilde\CL$. (The result is in continuous space, but the arguments work just as well for difference equations.) We refer to the solution for $\tilde\CL$ as $\tilde\pi_\beta$ and $\tilde\pi_\alpha$.

Using the trivial upper bound on $\tilde\pi_\beta(k)$ for $k \geq \bin{y_0}$, we find
\begin{equation}\label{bound}
\sum_{k \leq \bin{y_0}-1} \pi_\beta(k) \leq \sum_{k = \bin{x_0}+1}^{\bin{y_0}-1} \tilde\pi_\beta(k) + \sum_{k = \bin{y_0}}^{\bin{\kappa_a}-1} \frac{p_b(k)}{F_a(k)}.
\end{equation}
Notice that $\tilde\pi_\beta$ must equal $(F_a(k))^{-1}p_b(k)$ for $\bin{\tilde\kappa_b}+1 \leq k \leq \bin{x_0}$, as bids will not be queueing in those bins. Consequently, for the first term in the right-hand side of \eqref{bound} we have the bound
\[
\sum_{k = \bin{x_0}+1}^{\bin{y_0}-1} \tilde\pi_\beta(k) \leq 1 - \sum_{k=\bin{\tilde\kappa_b}+1}^{\bin{x_0}}\frac{p_b(k)}{F_a(k)} \leq 1 - \sum_{k = \bin{y_0}}^{\bin{\kappa_a}-1} \frac{p_b(k)}{F_a(k)} - \epsilon_0,
\]
as long as the bins are narrow enough. Indeed, notice that $x_0-\tilde\kappa_b > x_0 - \kappa_b = \kappa_a - y_0$ (from monotonicity of $\CL$ vs. $\tilde\CL$ and symmetry), the denominator is increasing in $k$, and the bid arrival density decreases with translation to the right. We require the bins to be narrow enough that the sums are all nonempty.\qedhere

{\bf Stage 1.} We now let $x_1, y_1$ be defined by $F_b(x_0) - F_b(x_1) = \epsilon_0 F_a(x_1)$ and $F_a(y_1) - F_a(y_0) = \epsilon_0 (1-F_b(y_1))$. Similarly to the argument for Stage 0, there exists a time $T_1$ such that for all $t \geq T_1$ there will be no fluid queues on $[\bin{x_1}+1,\bin{y_1}-1]$. Indeed, if there are fluid bids in the interval $[\bin{x_1}+1, \bin{x_0}]$, then whenever the highest bid is below $\bin{x_0}$ it is in fact in this interval; the defining inequality then means that the fluid amount of bids in this interval decreases, and similarly for asks.

Next, we use the difference equation description on $[\bin{x_1}+1, \bin{y_0}-1]$ to show that after $T_1$, the highest bid spends at least $\epsilon_1 > 0$ of its time below $x_1$. This will require comparison against a different restricted LOB $\tilde \CL_1$, where we merge all prices up to $\bin{x_1}+1$ and from $\bin{y_1}-1$.

{\bf Subsequent stages.} We can now construct a nested sequence of intervals $\dotsc < x_2 < x_1 < x_0 < y_0 < y_1 < y_2 < \dotsc$, where the inequalities are strict provided bins are narrow enough. It remains to show that $\lim_{k \to \infty, N \to \infty} x_k = \kappa_b$ and $\lim_{k \to \infty, N \to \infty} y_k = \kappa_a$. (Note that $N \to \infty$, i.e. thinner bins, is certainly necessary for this to hold!)

This result follows from the fact that $\epsilon_i$ can be taken to be bounded below:
\begin{equation}\label{joint density bound}
\epsilon_i \geq \sum_{k=\bin{\tilde\kappa_b}+1}^{\bin{x_i}}\frac{p_b(k)}{F_a(k)} - \sum_{k = \bin{y_i}}^{\bin{\kappa_a}-1} \frac{p_b(k)}{F_a(k)} \geq \left(\frac{1}{F_a(\bin{x_i})} - \frac{1}{F_a(\bin{y_i})}\right) \left(F_b(\bin{x_i}) - F_b(\bin{\tilde\kappa_b}+1) \right).
\end{equation}
As long as $x_i$ is bounded away from $\kappa_b$ (and bin widths are small enough), this will be bounded below, and therefore $x_i - x_{i+1}$ and $y_{i+1}-y_i$ will be bounded below.

{\bf Convergence to ODE.} The convergence of bounded solutions to difference equations to solutions of an ODE is standard. The argument above gives an inequality for the initial conditions, but note that as we approach $\kappa_b$ the initial conditions become exact. Indeed,
\[
F_a(\kappa_b+\epsilon)\varpi_b(\kappa_b+\epsilon) = \int_{\kappa_b+\epsilon}^1 \varpi_a(x) f_a(x) dx \to 1,
\]
since the lowest ask will never be below $\kappa_b$. Also,
\[
(F_a(x)\varpi_b(x))'\vert_{x=\kappa_b+\epsilon} = -\varpi_a(\kappa_b+\epsilon) = -(1-F_b(\kappa_b+\epsilon))^{-1}\int_0^{\kappa_b+\epsilon}\varpi_b(x) f_b(x) dx \to 0,
\]
since the highest bid density is bounded.\qedhere
\endproof

Putting this result together with Proposition~\ref{prop: weak distro} shows that, for symmetric distributions $p_b$, $p_a$ with $p_b$ decreasing, the fluid limits $\pi_\beta(k,t) / p_b(k)$, $\pi_\alpha(k,t) / p_a(k)$ will approach, as $t \to \infty$ and $N \to \infty$, the solution of the ODE \eqref{eq: ODE}, uniformly on compact subsets of $(\kappa_b, \kappa_a)$.

\begin{remark}
The argument leading to the inequality \eqref{joint density bound} implies that the joint density of the highest bid and lowest ask must be bounded away from zero on at least a fraction of the boundary of the support, i.e. the probability of the event ``there are no asks below $\kappa_a$ and the highest bid is at $\kappa_b + x$'' should be $O(x)$ but not $o(x)$. In fact, the simulated joint density in~\cite{EY_thesis} is bounded away from 0 everywhere except the very corner (highest bid at $\kappa_b$ and lowest ask at $\kappa_a$).
\end{remark}

It remains to show that stability of fluid limits implies positive recurrence of the Markov chain.

\begin{lemma}[Fluid stability and positive recurrence]\label{lm: fluid => recurrent}
Consider a LOB satisfying the assumptions of Theorem~\ref{thm: fluid stable}. Suppose that on some interval of bins $k_0 \leq k \leq k_1$, all fluid limits with initial state bounded above by $1$ satisfy the following: there exists a time $T$ (depending on $\{p_a(k), p_b(k), \text{bin widths}\}$), such that for all times $t \geq T$,
\[
q_b(k,t) = 0, \qquad q_a(k,t) = 0, \qquad k_0 \leq k \leq k_1,\ t \geq T.
\]

Consider a limit order book $\tilde\CL$ started with infinitely many bids
in bin $k_0-1$ and infinitely many asks in bin $k_1+1$; its state is
described by the Markov chain of queue sizes in bins $k_0 \leq k \leq k_1$.
The Markov chain associated with $\tilde\CL$ is positive recurrent.
\end{lemma}

\proof{Proof.}
To go between fluid stability and positive recurrence, we use
multiplicative Foster's criterion \cite[Theorem 13.0.1]{Meyn_Tweedie}.
Let
\[
Q(t) = \norm{(Q_b(k,t), Q_a(k,t))_{k_0 \leq k \leq k_1}},
\]
and let $C$ be sufficiently large. Let $Q(0) = q > C$, and consider the fluid scaling $\ov Q_{a,b}(k,t) = q^{-1}Q_{a,b}(k,qt)$. By Theorem~\ref{thm: fluid convergence}, if $C$ and hence $q$ is large enough, there exists a fluid limit $(q_a(k,t), q_b(k,t), \tau_\alpha(k,t), \tau_\beta(k,t))_{k_0 \leq k \leq k_1}$ satisfying $\norm{q_a(k,t), q_b(k,t)} = 1$, such that
\[
\BP(\norm{\ov Q_a(k,t) - q_a(k,t), \ov Q_b(k,t) - q_b(k,t)} > \epsilon) \leq \epsilon \quad \text{for all } t \in [0,T].
\]
In particular, $\BP(\norm{Q_a(k,qT), Q_a(k,qT)} > \epsilon q) < \epsilon$. Note further that $\norm{Q_a(k,qT), Q_b(k,qT)} \leq A(qT) + B(qT)$ is bounded by the arrival process, and hence has all moments. Thus, we conclude
\[
\BE_q[\norm{Q_a(k,qT), Q_b(k,qT)}] \leq \epsilon(1+2T)q.
\]
Choosing $\epsilon < (1+2T)^{-1}$ completes the proof.\qedhere
\endproof

\subsection{General order price distributions.}
It remains to remove the extra conditions (symmetric and decreasing) on the order price distributions, and finish the argument for continuous limit order books. This requires two observations:
\begin{enumerate}
\item Recall that a continuous LOB could be bounded by two discrete
LOBs with different arrival price distributions (in one of them,
we shift all arriving bids one bin to the left). This shifted arrival
distribution no longer satisfies the absolute continuity conditions, but
nevertheless, Lemma~\ref{lm: add one} shows that all of the above
fluid-scaled arguments work for it as bin size shrinks to 0. Specifically,
we model the bid arrivals as shifting the \emph{rightmost} bin of bids all
the way to the left, and then the difference between the two books is at
most two bins' worth of arrivals over the fluid time interval $[0,T]$, which will be small provided bins are narrow. This allows us to conclude the positive recurrence of a continuous LOB with infinitely many bids at price $\CP(\kappa_b)+\epsilon$ and infinitely many asks at price $\CP(\kappa_a)-\epsilon$, provided the densities $f_a$, $f_b$ are bounded above and below, symmetric, and $f_b$ is decreasing.
\item By Lemma~\ref{lm: decrease queues}, replacing the bid arrival price distribution by another distribution with stochastically higher prices, and/or replacing the ask arrival price distribution by another distribution with stochastically lower prices, results in fewer orders in a book. In particular, if we have shown the positive recurrence of an LOB with an infinite supply of bids at price $p$ and asks at price $q$ with a particular arrival distribution, the LOB will remain positive recurrent when we switch to an arrival price distribution with bids further right, and asks further left. Notice that as long as there are bids in the interval $(p,q)$, they evolve on that interval identically whether or not there is an infinite supply of bids at $p$; and similarly for asks. This can be used to show that fluid limits drain in the new LOB on the interval $(p,q)$.

In the new LOB with the shifted price distribution, $(p,q)$ may not be close to $(\tilde\kappa_b, \tilde\kappa_a)$, so we will be wanting to extend the interval, as in Claim~0.3 of Theorem~\ref{thm: fluid stable}. The argument there does not use the full extent of the symmetry and monotonicity conditions; they are only used to prove the inequality
\[
\sum_{k=\bin{\kappa_b}+1}^{\bin{p}}\frac{p_b(k)}{F_a(k)} \geq \sum_{k = \bin{q}}^{\bin{\kappa_a}-1} \frac{p_b(k)}{F_a(k)} + \epsilon
\]
for some $\epsilon > 0$. For this inequality to hold, it is entirely sufficient to have
\begin{equation}\label{eq: sufficient}
\int_{\hat\kappa_b}^p \frac{f_b(x)}{F_a(x)}dx \geq \int_q^{\tilde\kappa_a} \frac{f_b(x)}{F_a(x)}dx + \epsilon,
\end{equation}
with no constraints on what happens between $p$ and $q$.

Consequently, for a general pair of densities $(f_b, f_a)$ bounded below and above, we begin by finding $f_{b,0}, f_{a,0}$ with $F_{b,0} \geq F_b$, $F_{a,0} \leq F_a$ which are symmetric and for which $f_{b,0}$ is decreasing. (For example, we may take $f_{b,0} = f_{a,0} = \min(f_a, f_b)$ on most of the interval, with $f_{b,0}$ taking a large value near 0, and $f_{a,0}$ taking a large value near 1.) We use Theorem~\ref{thm: fluid stable} to show that fluid limits drain for $f_{b,0}, f_{a,0}$ (and hence, by Lemma~\ref{lm: decrease queues}, also for $(f_b, f_a)$) on an interval $(\kappa_{b,0}, \kappa_{a,0})$. We then modify $f_{b,1}$ on $(0, \kappa_{a,0})$ and $f_{a,1}$ on $(\kappa_{b,0},1)$ to find the next pair of bounded densities $(f_{b,1}, f_{a,1})$ for which $F_{b,0} \geq F_{b,1} \geq F_b$, $F_a \leq F_{a,1} \leq F_{a,0}$, and \eqref{eq: sufficient} holds. We already know from monotonicity that fluid limits will drain for these distributions on $(\kappa_{b,0}, \kappa_{a,0})$, and we use the inequality for $p \leq \kappa_{b,0}$ and $q \geq \kappa_{a,0}$ to extend fluid stability to the bigger interval $(\kappa_{b,1}, \kappa_{a,1})$. We repeat the process until the interval $(\kappa_{b,n}, \kappa_{a,n})$ approaches the entire interval $(\tilde\kappa_b, \tilde\kappa_a)$ for $(f_b, f_a)$.

To see that it will indeed approach the entire interval, notice that all that really matters for the thresholds of a LOB is $F_{a,b}(x), \kappa_b \leq x \leq \kappa_a$; it is immaterial what $f_b$ and $f_a$ do outside of those intervals, so long as they integrate to the correct amounts. Consequently, if $\kappa_{b,n} > \tilde\kappa_b + \epsilon$, it must be that $F_{b,n} < F_b$ or $F_{a,n} > F_a$ somewhere on $[\kappa_{b,n}, \kappa_{a,n}]$, which means that the process won't get ``stuck'' until $\kappa_{b,n} \searrow \tilde\kappa_b$ and $\kappa_{a,n} \nearrow \tilde\kappa_a$.
\end{enumerate}

\section{Discussion.}
\label{sec:applications}
In this section we discuss several applications of our
methods and results. We begin with a discussion of market orders
and then consider various simple trading strategies.

\subsection{Market orders.}

The orders we have considered so far, each with a price attached, are
called limit orders.  Suppose that, in addition to limit orders, 
there are also \emph{market orders} 
which request to be fulfilled immediately at
the best available price. Suppose that limit order bids and asks
arrive as independent Poisson processes of rates $\nu_b, \nu_a$ 
respectively; and that the prices associated with  limit order 
bids, respectively asks,
are independent identically distributed random variables with density 
$f_b(x)$, respectively $f_a(x)$.  Without loss of 
generality we may assume that $x \in (0,1)$. 
In addition suppose that there
are independent Poisson arrival streams of market order bids and asks
of  rates $\mu_b, \mu_a$ respectively. Then these correspond to 
extreme limit orders: we simply associate
a price $1$ or $0$ with a market bid or market ask respectively.

Note that, in addition to market orders, we have also allowed
an asymmetry in arrival rates between bid and ask orders. 
The intuition behind equations~\eqref{eq: integral_intro} leads
to the generalization
\begin{subequations}\label{eq: integral_intro-d}
\begin{equation} 
\nu_b f_b(x) \int_x^{\kappa_a} \pi_a(y) dy = \pi_b(x) \left(\mu_a + \nu_a
\int_{0}^x f_a(y)  dy \right)
\end{equation}
\begin{equation} 
\nu_a f_a(x) \int_{\kappa_b}^x \pi_b(y) dy = \pi_a(x)\left( \nu_b \int_x^{1}
f_b(y) dy + \mu_b \right) 
\end{equation}
\end{subequations}
although now the existence of a solution to these equations satisfying
the required boundary conditions is not assured,
and the deduction of the recurrence properties necessary for 
an interpretation of 
$ \pi_b(x), \pi_a(x)$ as limiting densities  may fail. 
To illustrate some of the possibilities we shall look 
in detail at a simple example.

Suppose $f_a(x)=f_b(x)=1, x \in (0,1)$, $\nu_a=\nu_b=1 -
\lambda$ and
$\mu_a=\mu_b=\lambda$. Thus a proportion $\lambda$ of all orders
are market orders. Use the notation $\pi_b(\lambda; x), 
\pi_a(\lambda; x)$ for 
the solution to
equations~\eqref{eq: integral_intro-d} satisfying
the required boundary conditions in this example. 
Then provided
$\lambda <w \approx 0.278$, the unique solution of $ w e^w = e^{-1}$,
this solution has $\pi_a(\lambda; x) = \pi_b(\lambda; 1-x)$ and
\begin{equation} \label{eq: soln-d}
\pi_b(\lambda; x) = \frac{1- \lambda}{1 +\lambda}  \cdot
\pi_b\left(\frac{1 + \lambda}{1-\lambda} x -
\frac{\lambda}{1- \lambda} \right), \quad x \in (\kappa(\lambda), 1-
\kappa(\lambda))
\end{equation}
where $\pi_b(\cdot)$ is the earlier solution~\eqref{eq: soln} and
\[
\kappa(\lambda)=  \frac{1 + \lambda}{1-\lambda} \cdot \frac {w}{1+w}  -
\frac{\lambda}{1- \lambda}.
\]
Indeed, provided $\lambda <w$ the
model is simply a rescaled version of the earlier model
with distribution~\eqref{eq: soln-d} having a support increased
from $(\kappa, 1-\kappa)$ to the wider interval
$(\kappa(\lambda), 1- \kappa(\lambda))$.
The inclusion of market orders in the model causes the 
price distributions to have atoms and not to be 
absolutely continuous with respect to each other; but nevertheless
the analysis of earlier sections continues to apply since the 
market orders arrive outside of the range $(\kappa(\lambda), 1-
\kappa(\lambda))$.

Next we explore this example as  
$ \lambda  \uparrow w$  and the support becomes the entire interval $(0,1)$.
In our model a market order bid,
respectively ask,  which
arrives when there are \emph{no} ask, respectively bid, limit orders in
the order book waits until it can be matched.  When $\lambda < w$ there
is a finite (random) time after which the order book
always contains limit orders of both types and no market orders
of either type and hence the analysis 
of previous sections applies. But if 
$ \lambda > w$ then infinitely often there will 
be no asks in the order book and infinitely often there will
be no bids in the order book, with probability 1. 
Now the difference between the number of bid and ask orders in the
limit book is a  simple symmetric random walk and hence null recurrent.
There will infinitely often be periods when the state of the order
book contains limit orders of both
types and no market orders
of either type, but such states cannot be positive recurrent.

In the model described above an arriving market order which cannot
be matched  immediately must wait until it can be matched. If instead
such orders are lost then we obtain a model which can be analyzed by 
the methods in Section~\ref{sec:marketmaking}: namely, we start the LOB with an infinite bid order at 0 and an infinite ask order at 1.

\subsubsection{Differing arrival rates.}\label{sec:differing} 

Our analysis in earlier sections assumed bids and asks
arrived at the same  rate. This was without loss of
essential generality, as it is convenient to
illustrate now with a discussion
of equations~\eqref{eq: integral_intro-d} when
$f_a(x)=f_b(x)=1, x \in (0,1)$,
$\nu_a, \nu_b >0$ and $\mu_a=\mu_b=0$. The solution to
equations~\eqref{eq: integral_intro-d} satisfying the
required boundary conditions is then
\begin{equation} \label{eq:darkappas}
\pi_b(x) = \kappa_a \left(\frac1x +\log\left(\frac{1-x}{x}\right)
-\frac{1}{\kappa_a} - \log\left(\frac{1-\kappa_a}{\kappa_a}\right)
\right),  \quad x \in (\kappa_b, \kappa_a)
\end{equation}
where
\begin{equation} \label{eq:kappas}
\nu_a \kappa_a = \nu_b ( 1- \kappa_b)
\end{equation}
and $\kappa_b$ is the unique solution to
\[
 \log\left(\frac{ (1- \kappa_b)^2 }{ \kappa_b (\nu_a /  \nu_b -1 +
\kappa_b)} \right)
=
\left( 1 + \frac{\nu_a}{ \nu_b} \right) \frac{1}{1- \kappa_b} .
\]
Although $\nu_a$ and $\nu_b$ may differ,
provided they are both positive 
the thresholds $\kappa_a $ and $ \kappa_b$ are both 
inside the interval $(0,1)$ and ensure
the necessary balance~\eqref{eq:kappas} between bids and asks 
that are matched.

If there are market orders, that is if
$\mu_a,\mu_b \geq 0$, then  this
results in a rescaling of the distribution~\eqref{eq:darkappas}
provided the support of the rescaled distribution 
remains contained within the interval $(0,1)$.

\subsubsection{Market impact.}  \label{sec:market_impact}

As a further illustration, consider the case where
$f_a(x)=f_b(x)=1, x \in (0,1)$,
$\nu_a = \nu_b =1$ and $\mu_a,\mu_b \geq 0$. 
Use the notation  $\pi_b(\mu_a,\mu_b; x), 
\pi_a(\mu_a,\mu_b; x)$ for
the solution to
equations~\eqref{eq: integral_intro-d} satisfying
the required boundary conditions in this case. 
Then provided
$\mu_a/(1+\mu_b), \mu_b/(1+\mu_a) <w ( \approx 0.278$) 
this solution has $\pi_a(\mu_a,\mu_b; x) = \pi_b(\mu_b,\mu_a; 1-x)$ and
\begin{equation} \label{eq: soln-m}
\pi_b(\mu_a,\mu_b; x) = 
\frac{\pi_b\left( (1 + \mu_b) x + \mu_a (1-x) \right)}{1+\mu_a+\mu_b},
\quad x \in \left( \frac{w (1 + \mu_b) - \mu_a}{w+1},
1- \frac{w (1 + \mu_a) - \mu_b}{w+1} \right)
\end{equation}
where $\pi_b(.)$ is the earlier solution~\eqref{eq: soln}.

An important assumption for  our mathematical development 
has been that all orders are for a single unit, and an outstanding
question concerns the extent to which the model can
be generalized. In practice, a long-term investor who wishes
to buy or sell a large number of units may choose to spread the order
in line with volume in the market, so as not to unduly move
the price against her~\cite{ELO}.  We are able
to analyze the market impact of
a particularly simple approach, when the investor
leaks the order into
the market according to an independent Poisson process over a 
relatively long period, where
 the market relaxes to the new equilibrium dynamics 
over that period.
Thus the impact of a large market order to buy 
will be to increase the parameter $\mu_b$ to say $\mu_b + \epsilon$. 
As $\epsilon$ increases the time taken to 
complete the order decreases, but the impact on the
distribution~\eqref{eq: soln-m} increases, leading to an overall less advantageous trading price.
Similarly if a large limit order is leaked into the market
as an independent Poisson process, this can also modeled by
a perturbation of 
equations~\eqref{eq: integral_intro-d}.

In markets with a relatively
small set of participants with large orders there may be advantages
in market designs where large transactions may be quickly arranged
at fixed prices; \cite{DZ} discuss trading protocols
that
complement limit order books for large strategic investors.

\subsubsection{One-sided markets.}  \label{sec:one-sided}

Toke~\cite{TOK} has considered a special case where
analytic
expressions for various quantities such as the 
expected number of bids in a given interval 
are readily available, as we now describe.

Suppose that 
$f_b(x)=1, x \in (0,1)$,
$\nu_a = \mu_b=0$ and $\mu_a > \nu_b >0$. Thus all bids
are limit orders and all asks are market orders, a \emph{one-sided market}. 
Then $\pi_b(x) =\nu_b/\mu_a, 
x \in (\kappa_b, 1)$ where $\kappa_b = 1 - \mu_a/\nu_b$. And, further, 
for $x > \kappa_b$ the 
number of bids present in the interval $(x,1)$, that is $B(x,1)$,  is a birth
and death process whose stationary distribution is geometric with mean
$\nu_b (1-x) / (\mu_a -\nu_b (1-x))$. 
Thus, for example, $\BE[B(x,y)]$ can be readily calculated. 

Various generalizations are also tractable, provided the market
remains one-sided~\cite{TOK}. For example, suppose each bid entering 
the LOB is cancelled after an independent exponentially distributed time with
parameter $\theta$ unless it has been previously matched. Then 
the
number of bids present in the interval $(x,1)$ is again a birth
and death process. Now the
entire LOB is a positive recurrent Markov process, 
and it is straightforward to 
verify 
that, as $\theta \downarrow 0$, bids to the left 
of $\kappa_b$ are seldom matched 
and the stationary distribution of the
rightmost bid approaches $\pi_b(x) =\nu_b/\mu_a, x \in (\kappa_b, 1)$, 
as we would expect.

\subsection{Trading strategies.}  \label{sec:tradstrat} 

Next we 
consider a few simple strategies that can be analyzed using our model. For
simplicity, we present the results for the case when the bid and ask price
distributions are equal and  uniform on $(0,1)$, but the
analysis easily extends to other arrival distributions. The limiting densities of the rightmost bid and leftmost ask for this model
were determined in Corollary~\ref{coroll: uniform}.

\subsubsection{Market making.} \label{sec:marketmaking}
We begin by considering a single market maker who places an
infinite number of bid, respectively ask, orders at $p$, 
respectively $q=1-p $, where $\kappa_b < p<q < \kappa_a $.
Thus whenever $q$ is the lowest ask
price, the trader obtains all bids that arrive at prices above $q$, and
whenever $p$ is the highest bid price, she obtains all asks that arrive at
prices below $p$, making a profit of $q-p$ per bid--ask pair so acquired.
Call the orders placed by the trader \emph{artificial}, to distinguish
them from the \emph{natural} orders. 
The rate at which the trader is able to match her
orders is proportional to $p$ times the probability that 
the rightmost bid is exactly $p$.

Placing an
 infinite supply of bids at a level below $\kappa$ has no asymptotic effect on the
evolution of the LOB. For $p > \kappa \approx 0.218$ no ask is 
accepted at a price less than $p$, and there will be a positive
probability that the rightmost bid is exactly $p$ (i.e., there are no bids at prices above $p$). To find this probability, we consider the following alternative model $\tilde \CL$: there is an infinite supply of bids placed at 0, but the price equivalence function $\tilde \CP$ is constant on $[0,p]$. (Otherwise, the initial state, arrival processes, and price equivalence functions coincide in $\CL$ and $\tilde \CL$.) In $\tilde \CL$, the bids and asks above $p$ will interact just as in $\CL$, so the probability that the rightmost bid is the infinite order in $\CL$ is equal to the probability that the rightmost bid is at or below $p$ in $\tilde \CL$. Note that only finitely many of the bids placed at 0 will ever be fulfilled in $\tilde \CL$. By Lemma~\ref{lm: add one}, pathwise, at all times the difference between the bid/ask queue sizes in $\tilde \CL$ and a limit order book without the infinite supply of bids at 0 will be bounded by the overall number of bids departing from that infinite supply. Hence the infinite bid at 0 is irrelevant for the analysis of the steady-state distribution of the highest bid, since in the limit $t \to \infty$ the difference will disappear.

In $\tilde \CL$, asks at prices below $p$ cannot stay in the book, i.e. $\varpi_a(x) = 0$ for $x \leq p$. By Remark~\ref{rem: density 1/x}, the density of the highest bid $\varpi_b(x)$ is equal to $1/x$ on $[\kappa_b, p)$, and to $C\left(\frac1x + \log\frac{1-x}{x}\right)$
on $[p, q]$ (the latter is obtained as in Corollary~\ref{coroll: uniform}). Recall that $\varpi_b$ is continuous, which allows us to determine $C$ and $\kappa_b$ (since $\varpi_b$ integrates to 1).
This allows us to find $\kappa_b$  as 
$$\kappa_b =
\frac{p}{e} \left(\frac{1-p}{p}\right)^C$$
and to deduce that
the rightmost natural bid has density
\[
\varpi_b(x) = \begin{cases}
\displaystyle { \frac1x} ,  &   \displaystyle {  \frac{p}{e} \left(\frac{1-p}{p}\right)^C
\leq x \leq p};\\
\displaystyle { C\left(\frac1x + \log\frac{1-x}{x}\right)}, & p  \leq x \leq
q;
\end{cases} 
\]
where $C = ({1+p\log((1-p)/p)})^{-1}.$
The probability the rightmost natural bid  is $p$ or less is
thus $1-C\log((1-p)/p)$, and this is therefore the probability that the
rightmost  bid  is exactly $p$ in the
model with infinitely many artificial bids at $p$ and 
infinitely many artificial asks at $q=1-p$, where $\kappa<p <1/2<q.$ 

To maximize the profit rate we need to solve  the optimization problem
\[
\begin{aligned}
&\text{maximize} && (1-2p)p\left(1-C\log\frac{1-p}{p}\right)\
&\text{where} && C = \left(1+p\log\frac{1-p}{p}\right)^{-1}\\
&\text{subject to} && p \in [\kappa,1/2].
\end{aligned}
\]
The maximum is attained at $p \approx 0.377$, and gives a profit 
rate of $\approx 0.054$.

\subsubsection{Sniping.} \label{sec:sniping}
We next consider a trader with a  sniping strategy: the trader immediately 
buys every bid that joins the LOB at price above $q$, and every ask
that joins the LOB  at price below $p$ (with $q=1-p$ still). Now the
trader has lower priority than the orders already in the queue, but she
obtains a better price for the orders that she does manage to buy. 

The effect on the LOB of the sniping 
strategy is to ensure there are no queued bids above
$q$ and no queued asks below $p$; for $p < q$, the set of bids and asks on
$(p,q)$ has the same distribution 
in the sniping and the market making model, and therefore the probability that $p$ is the highest bid is the same as in the market making model as well. (The profit rates for the trader are different.) But it also makes sense to consider the
sniping strategy with $p > q$, when it ensures that there are no queued orders of any kind in the interval $(q,p)$: they are all sniped up by the trader. (An ask arriving at price $a \in (q,p)$ cannot be matched with a queued bid, because there are no queued bids above $q$.) The trader makes a net profit of zero on the orders in $(q,p)$; the point of sniping them is to increase the probability of being able to buy a bid at a high price.

Summarizing, if $p > q$ 
then the LOB has no queued orders between $p$ and $q$. Since all the bids are at prices below $q$, and the ask density there is zero, we seefrom Proposition~\ref{prop: weak distro} that the density of the rightmost bid is $\varpi_b(x) = 1/x$ on $[\kappa_b,q)$; since $\varpi_b$ integrates to 1, we find $\kappa_b = q/e$. 
Notice that the distribution of the rightmost bid 
 stochastically decreases as $q$ decreases, hence the probability of acquiring an ask at low price $a < 1/2$ increases as $q$ decreases. This shows that 
the profit rate from sniping bids above $q$ and asks below $p$ for $p >
1/2$ is strictly higher than the profit rate 
from sniping bids above $p$ and asks below $q$. Thus, it suffices to consider the case of $p > 1/2 > q$. We thus solve
\[
\begin{aligned}
&\text{maximize} && \int_{\kappa_b}^{1-p}(1-2x)\log\frac{x}{\kappa_b}dx\
&\text{where} && \kappa_b = \frac{1-p}{e} \\
&\text{subject to} && p \in [1/2,1].
\end{aligned}
\]
The maximum is attained at $1-p= q =e/(e^2+1) \approx 0.324$ and gives a
profit rate of $\approx 0.060$.

\begin{figure}
\begin{center}
{\includegraphics[width=0.5\textwidth]{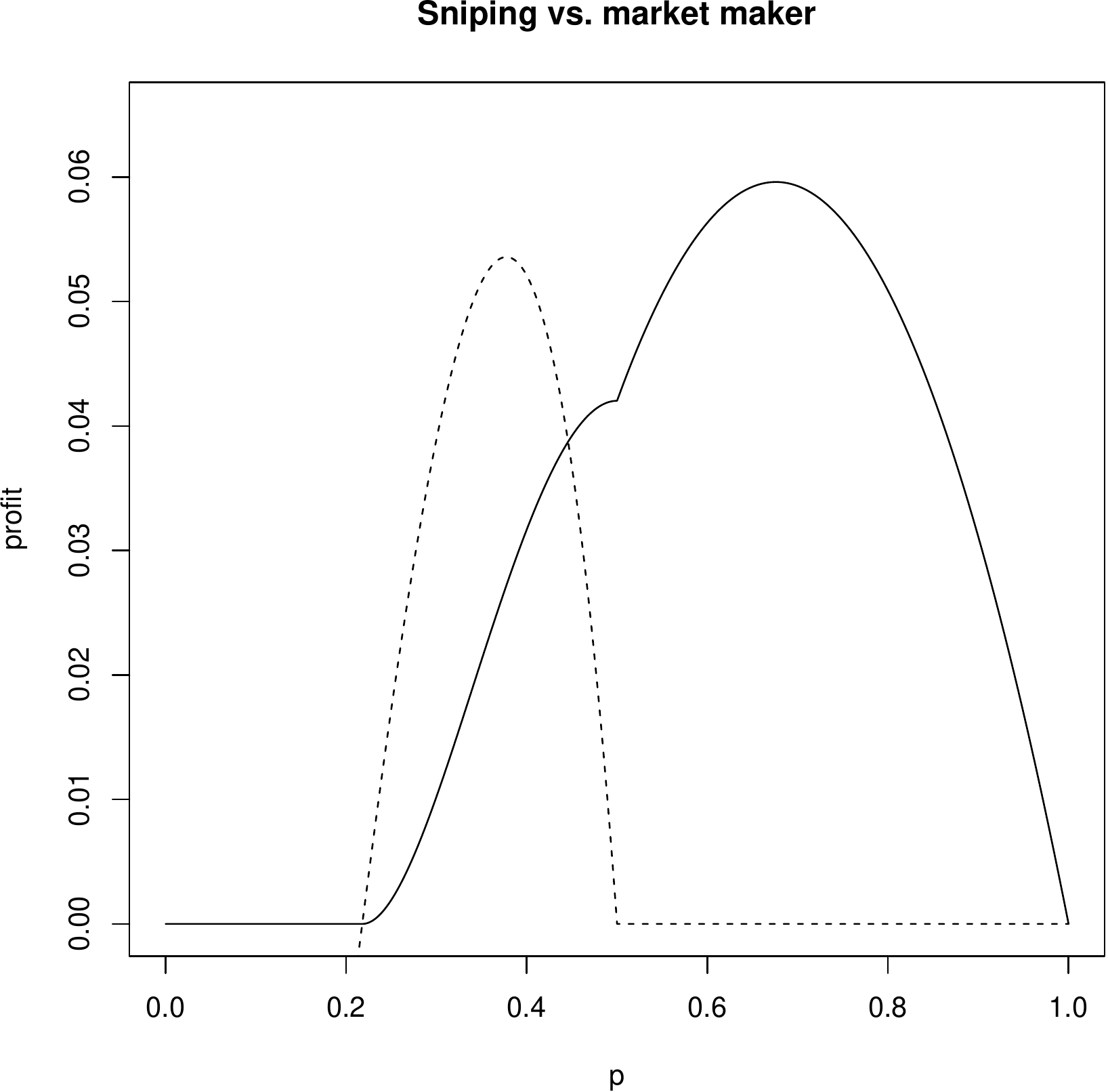}}
\end{center}
\caption{Profit from sniping and market making strategies. Solid line is the sniping strategy, dashed line is the market making strategy. (Sniping with $p < 1/2$ is shown for completeness; as argued in the text, it does not maximize the profit.)}
\label{fig: sniping and sluggish}
\end{figure}

Figure~\ref{fig: sniping and sluggish} presents a comparison between the 
profit rates from the market making and sniping strategies, as a function
of $p$ (which, recall, is the price below which the trader would like all
asks) -- for completeness, $p < 1/2$ is included for the sniping strategy as well.

\subsubsection{A mixed strategy.}  \label{sec:mixedstrategy} 
It is possible to consider a mixture of the above strategies: 
the trader places an infinite supply of bids at $P$ (thus acquiring all asks that arrive below $P$ whenever $P$ is the highest bid price), but in addition attempts to snipe up all the additional asks that land at prices $x < p$. We assume the trader gets the best of the two possible prices when both $p$ and $P$ are larger than the price of the arriving ask. There are several possible cases corresponding to the relative arrangement of $p$, $P$, and $1/2$:
\begin{enumerate}
\item If $p < P$ (this means that there are no additional asks to snipe
up), this degenerates to the market maker strategy, with a profit of
$(1-2P)$ per bid--ask pair bought, with pairs bought at rate
$P\log(P/\kappa_b)$. (The probability of the highest natural bid being
below $P$ is $\log(P/\kappa_b)$; when it is there, asks arrive at prices below $P$ at rate $P$.) Clearly, one wants $P < 1/2$ in this case, otherwise the profit is negative, so we can write this case as $p < P < 1/2$.
\item If $P < p < 1/2$, then one gets additional asks at price $x$ at rate $\log(x/\kappa_b)$, for a profit of $(1-2x)$, for all $x$ from $P$ to $p$.
\item If $P < 1/2 < p$, there are two further cases: we may have $P < 1-p$ or $P > 1-p$.
\begin{enumerate}
\item If $P < 1-p < 1/2 < p < 1-P$, then the trader snipes all orders between $1-p$ and $p$ for a net profit of 0. Profit $(1-2P)$ from a bid--ask pair matching the infinite orders is generated at rate $P\log((1-p)/\kappa_b)$, and profit $1-2x$, $P \leq x \leq 1-p$, from sniping is generated at rate $1+\log(x/\kappa_b)$. By Remark~\ref{rem: density 1/x}, the highest bid density is $1/x$ on $(\kappa_b,1-p]$, so $\kappa_b = (1-p)/e$.
\item If $1-p < P < 1/2 < 1-P < p$, then $P$ is always the best bid, which means that the trader gets all the asks that arrive below $P$, generating profit at rate $(1-2P)P$. Orders arriving between $P$ and $1-P$ cancel each other, and all the asks arriving between $1-P$ and $p$ are bought up for a loss (negative profit) of $(1-2x)$.
\end{enumerate}
\item Finally, the case $P > 1/2$ is silly, because every bid--ask pair bought will be bought at a loss.
\end{enumerate}

\begin{figure}
  \begin{center}
    {\includegraphics[width=0.5\textwidth]{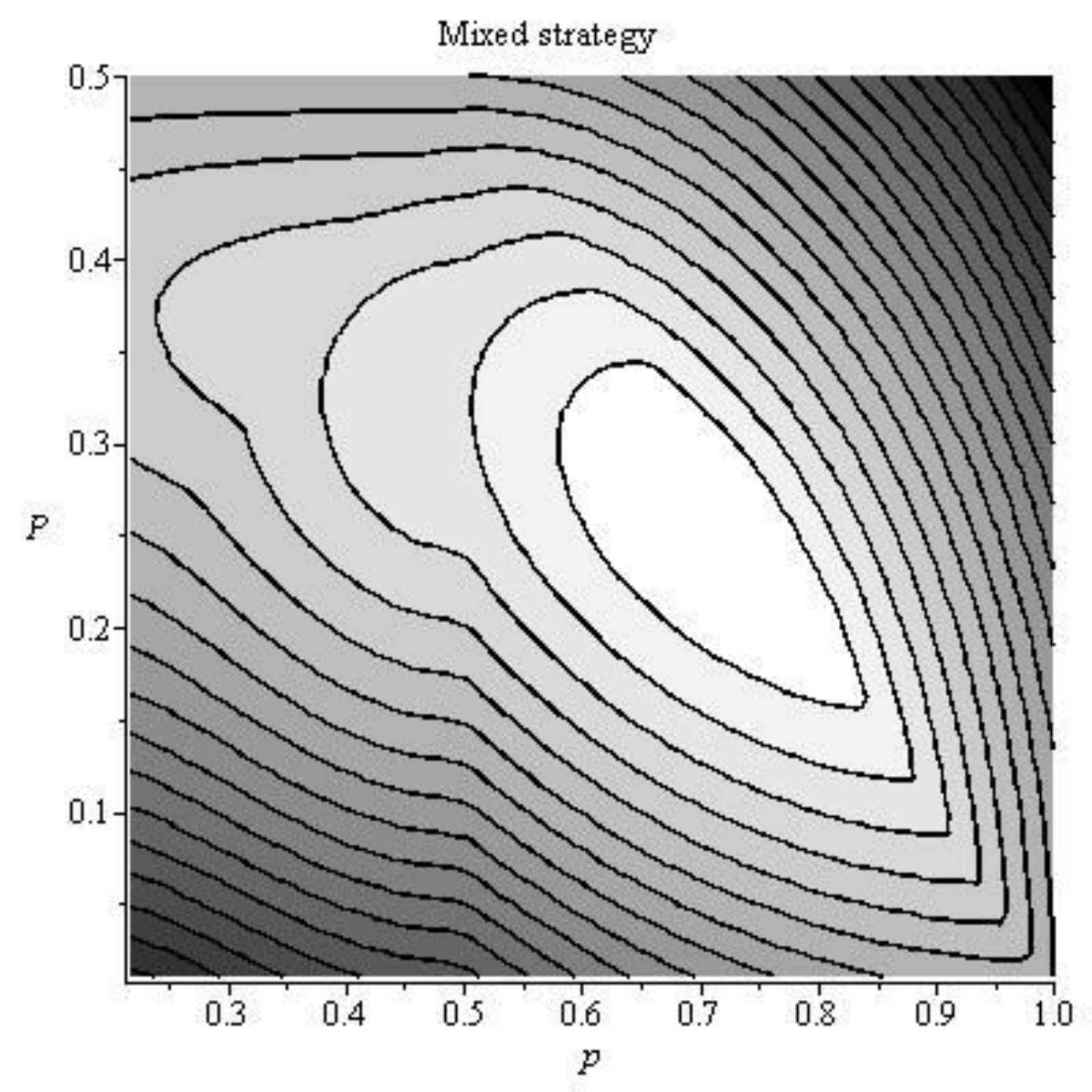}}
    \end{center}
  \caption{Profit rate from the mixed strategy as a function of sniping threshold $p$ and infinite bid order location $P$.}
  \label{fig: mixed model}
\end{figure}

Figure~\ref{fig: mixed model} shows the profit for the two-parameter space. The largest profit is obtained when $P = 1-p = 1/4$, and the profit is then acquired at rate $1/8 = 0.125$. This corresponds to the trader placing an infinite bid order at $1/4$ (thus buying all asks that arrive with price below $1/4$ for $1/4$), an infinite ask order at $3/4$, and sniping up all orders that join the LOB at prices between $1/4$ and $3/4$. 

\subsection{Competition between traders.}

Finally we  comment on 
the situation that arises when multiple traders
compete using the simple strategies described
in Section~\ref{sec:tradstrat}.

Consider first the case of two competing traders, the
first of whom has the ability to employ a sniping strategy
of the form described in Section~\ref{sec:sniping},
and the second of whom cannot act quickly enough to snipe
but does have the capacity to 
employ a market making strategy of the form described in
Section~\ref{sec:marketmaking}. Suppose then the market maker
places an infinite number of bid, respectively ask, orders
at $P$, respectively $1-P$, where $P \leq 1/2$. And suppose 
the sniper immediately buys every bid that joins the LOB
at price above $q$, and every ask that joins the LOB
at price below $1-q$, where $P \leq q \leq 1/2$.  

For given $P$ and $q$ the behavior of the LOB is as
analyzed in Section~\ref{sec:mixedstrategy}: but incentives
for the two traders are different. Given $P$ and $q$
the profit rate for the sniper is 
$\int_P^q (1-2x) \ln(ex/q) dx$
and for the market maker $(1-2P)P \int_{q/e}^P 1/x \ dx = (1-2P)P\log(eP/q)$
provided $P > \kappa_b = q/e$. If $P<q/e$ the market maker's orders
are outside of the recurrent range $(\kappa_b, 1- \kappa_b)$ of
the LOB and so are not matched.  
It is natural to suppose the sniper follows the market maker: that is 
the sniper
observes the choice $P$ of the market maker and chooses $q$
accordingly. The maximizing choice for the sniper is 
then 
$q = \sqrt{P(1-P)}$.
Given this, the optimum choice for 
the market maker, that maximizes his profit rate, is 
$P \approx 0.340$.
At this equilibrium, the profit rate of the market maker is 
0.073
and of the sniper 
0.020.

Consider next the case of two or more traders using either the 
market making strategies of Section~\ref{sec:marketmaking} or the 
mixed strategies of Section~\ref{sec:mixedstrategy}. 
Each trader will have an incentive to
improve slightly the prices at which  she places infinite orders
of bids and asks, thus gaining all the
profit from those orders for herself alone. 
The Nash equilibrium has the traders compete away the bid-ask spread
and with it all their profits. The model becomes an example of
the 
Bertrand model of price competition and, as there, the conclusion is softened
with more realistic assumptions on, for example,
capacity constraints or cost asymmetries.

Next consider competition between two or more traders using the 
sniping strategies of Section~\ref{sec:sniping}; for example
between traders who 
attempt to snipe a limit order as it arrives with an exactly
matching limit order. 
If multiple traders attempt to simultaneously snipe 
the arriving order, 
than one
of them will succeed and the others will 
cancel their own orders immediately as they detect that their orders
have not been successful. 
There is clearly an 
advantage for a trader who can snipe an arriving order more quickly than the
other traders, and indeed such a trader can enforce the 
optimum sniping strategy of Section~\ref{sec:sniping} and exclude slower traders
from the market. 
It has been argued that competition on speed
is wasteful (see \cite{BCS}), and there are  proposals to 
encourage traders to 
compete on price, rather than speed, as for example in the
proposal of \cite{BCSAER} where   a
market continuous in time is replaced with frequent batch auctions, held perhaps
several times a second. We shall explore the consequences of competition
on price between sniping traders who can all react at the
same speed to a new order entering the LOB. 

In such a competitive environment traders will have an incentive
to increase the price $q$ above which they snipe bids, and decrease
the price $p$ below which they snipe asks, towards $1/2$: they
will refrain from sniping orders on
which they would expect to make a loss. At the Nash equilibrium, 
each trader will snipe at all asks with prices below $1/2$ and at 
all bids with prices
above $1/2$ 
(getting the order with the same probability
as each of the other traders). 
The rightmost bid will then have density $1/x$ on 
$(1/(2 e), 1/2)$ by Remark~\ref{rem: density 1/x}. This results in a combined profit  rate 
$ 1/(2 e) - (1+e^2)/(8e^2) \approx 0.042 $. 
Thus price competition between sniping 
traders has decreased their combined profit rate only slightly, 
from $0.060$ to $ 0.042 $. Alternatively, one can view this
reduction as the effect of a batch rather than a continuous market. 


Next we comment on the impact of traders on the bid-ask spread. 
The mean of the distribution~\eqref{eq: soln} can be 
calculated and is simply $(1-\kappa)/2$. Thus 
without traders the mean spread between the highest bid and the
lowest
ask in the LOB is $\kappa \approx 0.218$, while the maximum spread is
$1-2\kappa \approx 0.564$.  
At the Nash equilibrium between sniping traders both are increased, 
the mean spread to $1/e \approx 0.368$
  and the maximum spread to $1- 1/e \approx 0.632$. For comparison,
with a single sniping trader both are further increased, the
mean spread to $ 1- 2(e-1)/(e^2+1) \approx 0.590$
  and the maximum spread to $ 1- 2/(e^2+1) \approx 0.762$; and at the equilibrium between a market making trader and a sniping trader the mean and maximum spread are as low as 0.228 and 0.320 respectively. These
calculations are of course for a specific example, but they do
illustrate the tractability of the model and its insights. 


As a final remark we comment on the inventory of traders under the Nash
equilibrium between sniping traders
described above. Observe that  
the LOB below $1/2$ evolves independently of the LOB above $1/2$,
and both processes are positive recurrent inside their corresponding thresholds. Consider the net position of the traders 
collectively, that is all the bids they have matched minus all the asks they have matched,
observed at those times when the LOB is empty. This evolves as a symmetric
random walk, and is null recurrent. Slight variations of the traders'
 strategies would moderate this conclusion: for example, a trader might
refrain from sniping bids close enough to $1/2$ when his net position is large. And of course such variations will be essential over longer time-scales than those considered in this paper where the arrival price distributions may vary.

%
%
%

\section*{Acknowledgments.}
The authors are grateful to Darrell Duffie for valuable early
comments on this work, and to the associate editor and the referees for
their careful
reviews.  The authors thank Jan Swart for drawing 
their attention to the references~\cite{Luckock, Plackova, Stigler} and his
preprint~\cite{Swart}, and for pointing out a mistake in an earlier statement of part 1 of Theorem~\ref{thm: threshold}. (Using different methods 
\cite{Swart} proves and gives an extensive discussion
of  part 3 of Theorem~\ref{thm: threshold}, while 
part 2
of  Theorem~\ref{thm: threshold} 
solves open problems from~\cite{Swart}.)
The second author's research was partially supported by NSF Graduate Research Fellowship and NSF grant DMS-1204311.

\bibliographystyle{abbrv}
\bibliography{lob}

\begin{thebibliography}{10}

\bibitem{AW}
I.~Adan and G.~Weiss.
\newblock Exact {FCFS} matching rates for two infinite multitype sequences.
\newblock {\em Operations Research}, 60:475--489, 2012.

\bibitem{Bramson}
M.~Bramson.
\newblock {\em Stability and Heavy Traffic Limits for Queueing Networks: St.
  Flour Lectures Notes}.
\newblock Springer, 2006.
\newblock \url{http://www.math.duke.edu/~rtd/CPSS2007/Bramson.pdf}.

\bibitem{BCS}
E.~Budish, P.~Cramton, and J.~Shim.
\newblock The high-frequency trading arms race: Frequent batch auctions as a
  market design response.
\newblock
  \url{http://faculty.chicagobooth.edu/eric.budish/research/HFT-FrequentBatchAuctions.pdf},
  2013.

\bibitem{BCSAER}
E.~Budish, P.~Cramton, and J.~Shim.
\newblock Implementation details for frequent batch auctions: Slowing down
  markets to the blink of an eye.
\newblock {\em American Economic Review}, 104:418--–424, 2014.

\bibitem{Cont_deLarrard}
R.~Cont and A.~de~Larrard.
\newblock Price dynamics in a markovian limit order book market.
\newblock {\em SIAM Journal of Financial Mathematics}, 4:1--25, 2013.

\bibitem{CST}
R.~Cont, S.~Stoikov, and R.~Talreja.
\newblock A stochastic model for order book dynamics.
\newblock {\em Operations Research}, 58:549–--563, 2010.

\bibitem{DZ}
D.~Duffie and H.~Zhu.
\newblock Size discovery.
\newblock NBER Working Paper No. 21696, 2015.
\newblock \url{http://dx.doi.org/10.3386/w21696}.

\bibitem{ELO}
D.~Easley, M.~L. de~Prado, and M.~O'Hara.
\newblock The volume clock: Insights into the high frequency paradigm.
\newblock {\em The Journal of Portfolio Management}, 39:19--29, 2012.

\bibitem{Gamarnik_Katz}
D.~Gamarnik and D.~Katz.
\newblock The stability of the deterministic {S}korokhod problem is
  undecidable.
\newblock {\em Queueing Systems}, pages 1--29, 2014.

\bibitem{GDDD}
X.~Gao, J.~G. Dai, A.~B. Dieker, and S.~J. Deng.
\newblock Hydrodynamic limit of order book dynamics.
\newblock \url{http://arxiv.org/pdf/1411.7502.pdf}, 2014.

\bibitem{GPSMFD}
M.~D. Gould, M.~A. Porter, S.~Williams, M.~McDonald, D.~J. Fenn, and S.~D.
  Howison.
\newblock Limit order books.
\newblock {\em Quantitative Finance}, 13:1709--1742, 2013.

\bibitem{KEN}
D.~Kendall.
\newblock Some problems in the theory of queues.
\newblock {\em Journal of the Royal Statistical Society}, 13(2):151--185, 1951.

\bibitem{LLLL}
A.~Lachapelle, J.-M. Lasry, C.-A. Lehalle, and P.-L. Lions.
\newblock Efficiency of the price formation process in presence of high
  frequency participants: a mean field game analysis.
\newblock \url{http://arxiv.org/pdf/1305.6323v3.pdf}, 2014.

\bibitem{LRS}
P.~Lakner, J.~Reed, and F.~Simatos.
\newblock Scaling limit of a limit order book model via the regenerative
  characterization of {L}{\'e}vy trees.
\newblock arXiv preprint \url{arXiv:1312.2340}, 2013.

\bibitem{LRSt}
P.~Lakner, J.~Reed, and S.~Stoikov.
\newblock High frequency asymptotics for the limit order book.
\newblock Submitted, 2013.
\newblock \url{http://people.stern.nyu.edu/jreed/Papers/LimitFinal.pdf}.

\bibitem{Luckock}
H.~Luckock.
\newblock A steady-state model of the continuous double auction.
\newblock {\em Quantitative Finance}, 3(5):385--404, 2003.

\bibitem{MMZ}
C.~Maglaras, C.~C. Moallemi, and H.~Zheng.
\newblock Queueing dynamics and state space collapse in fragmented limit order
  book markets.
\newblock Columbia Business School Research Paper No. 14-13, 2014.
\newblock \url{http://dx.doi.org/10.2139/ssrn.2403884}.

\bibitem{Meyn_Tweedie}
S.~Meyn and R.~L. Tweedie.
\newblock {\em Markov Chains and Stochastic Stability}.
\newblock Cambridge University Press, second edition, 2009.

\bibitem{Plackova}
J.~Pla{\v c}kov{\'a}.
\newblock {\em Shluky volatility a dynamika popt{\'a}vky a nab{\'i}dky}.
\newblock Master Thesis, MFF, Charles University Prague, 2011.
\newblock (In Czech).

\bibitem{Rosu}
I.~Ro{\c{s}}u.
\newblock A dynamic model of the limit order book.
\newblock {\em Review of Financial Studies}, 22:4601--4641, 2009.

\bibitem{SIM}
F.~Simatos.
\newblock Coupling limit order books and branching random walks.
\newblock {\em Journal of Applied Probability}, 51:625--639, 2014.

\bibitem{Stigler}
G.~Stigler.
\newblock Public regulation of the securities markets.
\newblock {\em The Journal of Business}, 37(2):117--142, 1964.

\bibitem{SY}
A.~L. Stolyar and E.~Yudovina.
\newblock Systems with large flexible server pools: Instability of
  {``}natural{''} load balancing.
\newblock {\em Annals of Applied Probability}, 23:2099--2138, 2013.

\bibitem{Swart}
J.~Swart.
\newblock Rigorous results for the stigler-luckock model for the evolution of
  an order book.
\newblock Submitted, 2016.
\newblock \url{arXiv:1605.01551}.

\bibitem{TOK}
M.~Toke.
\newblock The order book as a queueing system: average depth and influence of
  the size of limit orders.
\newblock {\em Quantitative Finance}, 14, 2014.

\bibitem{EY_thesis}
E.~Yudovina.
\newblock {\em Collaborating Queues: Large Service Network and a Limit Order
  Book}.
\newblock Ph.D. thesis, University of Cambridge, 2012.

\bibitem{ZCW}
S.~A. Zenios, G.~M. Chertow, and L.~M. Wein.
\newblock Dynamic allocation of kidneys to candidates on the transplant waiting
  list.
\newblock {\em Operations Research}, 48:549--569, 2000.

\end{thebibliography}

\end{document}